\documentclass[12pt]{article}
\usepackage{amssymb,amsmath}
\usepackage{latexsym}
\usepackage{graphicx}
\usepackage{subfigure}
\usepackage{booktabs}
\usepackage{bbm}
\usepackage{enumitem}
\usepackage{color}
\usepackage{datetime}
\usepackage{cite}
\usepackage[margin=20pt,small]{caption}

\usepackage{psfrag}

\usepackage[
      colorlinks=false,
      linkcolor=darkblue,  
      urlcolor=blue,    
      filecolor=blue,     
      citecolor=red,
linktocpage=true,
      pdfstartview=FitV,
      bookmarksopen=true    
      ]{hyperref}

\let\oldbibliography\thebibliography
\renewcommand{\thebibliography}[1]{\oldbibliography{#1}
\setlength{\itemsep}{0pt}} 

\numberwithin{equation}{section}  

\makeatletter
\g@addto@macro\bfseries{\boldmath}
\makeatother

\definecolor{cardinal}{rgb}{0.6,0,0}
\definecolor{darkgreen}{rgb}{0,0.5,0}
\definecolor{golden}{rgb}{0.92, 0.7, 0}
\definecolor{midnight}{rgb}{0, 0, 0.5}
\definecolor{darkblue}{rgb}{0.2, 0, 0.8}
\newcommand{\Red}{\color{red}}

\def\IC{\mathbb{C}}
\def\IP{\mathbb{P}}
\def\Neql#1{{\cal N}\!=\!{#1}}
\def\coeff#1#2{\relax{\textstyle {#1 \over #2}}\displaystyle}

\def\IR{\mathbb{R}}
\def\ZZ{\mathbb{Z}}

\def\cB{{\cal B}}

\def\cI{{\cal I}}

\def\cK{{\cal K}}
\def\cL{{\cal L}}
\def\cM{{\cal M}}
\def\cN{{\cal N}}
\def\cP{{\cal P}}

\def\cR{{\cal R}}

\def\cV{{\cal V}}
\def\nBPS#1{$\frac{1}{#1}$-BPS}

\def\cO{{\cal O}}

\def\cmnt#1{{\Red \noindent [[KP: \,#1\,]]}}

\def\suthuu{ SU(3)$\times$\rm U(1)$\times$U(1)}

\begin{document}

\begin{titlepage}

\bigskip
\bigskip
\centerline{\LARGE \bf  Flowing to Higher Dimensions:}
\medskip
\centerline{\LARGE \bf  A New Strongly-Coupled Phase on M2 Branes }

\bigskip\bigskip\bigskip

\centerline{{\bf Krzysztof Pilch, Alexander Tyukov   and Nicholas P. Warner}}
\bigskip

\centerline{ Department of Physics and Astronomy,}
\centerline{University of Southern California,} \centerline{Los
Angeles, CA 90089, USA}
\bigskip
\bigskip
\centerline{{\rm pilch@usc.edu,~tyukov@usc.edu,~warner@usc.edu} }
\bigskip
\bigskip

\bigskip\bigskip

\begin{abstract}
\noindent
We describe a one-parameter family of new  holographic RG flows that start from $AdS_4 \times S^7$  and go to  $\widehat{AdS_5} \times \cB_6$, where $\cB_6$ is conformal to a   K\"ahler manifold and $\widehat{AdS_5}$ is Poincar\'e $AdS_5$ with one spatial direction compactified and fibered over $\cB_6$.  The new solutions  ``flow up dimensions,'' going from the $(2+1)$-dimensional conformal field theory on M2 branes in the UV to a $(3+1)$-dimensional field theory on intersecting M5 branes in the infra-red. The M2 branes completely polarize into M5 branes along the flow and the Poincar\'e sections of the $\widehat{AdS_5}$ are the $(3+1)$-dimensional common intersection of the M5 branes.  The emergence of the extra dimension in the infra-red suggests a new strongly-coupled  phase of the M2 brane and ABJM theories in which charged solitons are becoming massless.  The flow solution is first analyzed by finding a four-dimensional $\Neql2$ supersymmetric flow in $\Neql8$  gauged supergravity. This is then generalized to a one parameter family of non-supersymmetric flows.  The infra-red  limit of the solutions appears to be quite singular in four dimensions but the uplift to eleven-dimensional supergravity is remarkable and regular (up to orbifolding).   Our construction is a  non-trivial application of the recently derived uplift formulae for fluxes,  going well beyond the earlier constructions of stationary points solutions. The eleven-dimensional supersymmetry is also analyzed and shows how, for the supersymmetric flow, the M2-brane supersymmetry in the UV is polarized entirely into M5-brane supersymmetry in the infra-red.
\end{abstract}

\end{titlepage}


\tableofcontents

\section{Introduction}
\label{Sect:introduction}

Brane polarization is one of the most remarkable and non-trivial aspects of string theory, arising from the fact that  branes couple to forms of all degrees \cite{Myers:1999ps}. In the supergravity limit, brane polarization comes about through the Chern-Simons terms:   sources of one type of brane can be generated through interactions of fluxes usually associated with other types of branes.    Such mechanisms have played a major role in understanding the possible infra-red phase structure of holographic renormalization group (RG) flows and in the fuzzball program in which singular sources of branes that would normally back-react into a black hole can be replaced by smooth cohomological fluxes and thereby generate a smooth, solitonic, microstates geometries. 

Within holographic field theory, one of the  earliest and most important examples of  brane polarization appeared in the  Polchinski-Strassler flow\cite{Polchinski:2000uf}  where the holographic duals of confining gauge theories were seen to require brane polarization of D3-branes  into 5-branes.  The early descriptions of two-charge fuzzballs did not really involve brane polarizations directly: such solutions  start with the D1-D5 electric charges and then the addition of dipolar KKM fields and angular momentum yields the smooth supertube backgrounds.  On the other hand, the three-charge problem and the entire microstate geometry program uses brane polarization in an essential way. This can be described in many duality frames (see, for example, \cite{Bena:2008dw}) but the simplest such frame is probably the original frame in which it was discovered \cite{Bena:2005va,Berglund:2005vb,Bena:2007kg}, where three distinct M2-brane electric charges polarize into cohomological fluxes associated with magnetic M5 branes.  Indeed, these microstate geometries have three M2-brane charges as seen from infinity, and look like a black hole from a distance, while the core of the solution  has three sets of dipolar M5-brane flux supported on homology cycles and no localized M2-brane sources.   Such a geometric transition has thus sometimes been described as M2-brane charges dissolved in M5-brane fluxes.  

One can see very explicitly how this can arise from the Maxwell equations of eleven-dimensional supergravity, which   may be written:\footnote{We are using ``old'' supergravity conventions.}
\begin{equation}
d*F^{(4)} ~+~  F^{(4)}  \wedge F^{(4)}  ~=~0 \,.
\label{Feom}
\end{equation}
This means that the {\it conserved}  M2  charge is not $\int *F$ but is  the ``Page charge''  defined by:
\begin{equation}
Q~=~ \int \,( *(F^{(4)} ) ~+~  A^{(3)}   \wedge F^{(4)}  )\,.
\label{PageChg}
\end{equation}
Thus, depending on the configuration, the M2 charge can come from electric ``M2 sources'' contributing via $*(F^{(4)} )$, or from magnetic ``M5 sources'' contributing via $A^{(3)}   \wedge F^{(4)} $, or through a combination of both.

The geometric transition between M2 and M5 brane sources has  played a number of roles in holographic field theory.  At the simplest level, if one starts with the field theory on a large stack of M2 branes then the holographic dual is simply $AdS_4 \times S^7$ and  turning on fermion masses or vevs of fermion bilinears necessarily means turning on magnetic M5-brane fluxes on $S^7$.  This has led to some very interesting holographic renormalization group flows.  In particular, one can turn on a particular set of fermion and boson mass terms and flow to conformal fixed points.  Such flows have been extensively studied using four-dimensional, $\Neql8$ gauged supergravity  \cite{Ahn:2000mf,Ahn:2001by,Ahn:2001kw,Ahn:2002eh,Ahn:2002qga,Bobev:2009ms,Bobev:2010ib} and involve turning on scalar fields in supergravity and moving between different critical points of the scalar potential in four dimensions \cite{Warner:1983du,Warner:1983vz}.  Uplifts to eleven dimensions for some of these solutions have been obtained and analyzed in \cite{Corrado:2001nv,Benna:2008zy}. 

There are also interesting classes of holographic flows in which, from the four-dimensional perspective,  the scalars flow to infinite values at the infra-red end of the flow.   While apparently singular, some of these flows have simple and physically-important interpretations that become apparent once one uplifts the solution to eleven dimensions.   One class of such flows are the Coulomb-branch flows for which, in the infra-red limit, the M2 branes spread out into some distribution in space  \cite{Cvetic:1999xx,Cvetic:2000zu,Gowdigere:2002uk} much as in the Coulomb-branch flows for D3 branes \cite{Kraus:1998hv,Freedman:1999gk,Gowdigere:2005wq}, which correspond to $\rm SU(N)$ gauge theory being broken, typically, to ${\rm U(1)}^N$.  

While the Coulomb-branch flows do not involve brane polarization, a class of very interesting  generalizations were studied in \cite{Pope:2003jp,Bena:2004jw}.   The crucial observation in \cite{Pope:2003jp} was that there was a manifest $\rm U(1)$ symmetry of a particular truncation of four-dimensional $\Neql8$ supergravity that related very different M-theory solutions and apparently very different holographic field theories.  This unexpected $\rm U(1)$  symmetry rotated bosonic bilinears into fermionic bilinears in the M2-brane field theory and rotated metric modes into flux modes in M-theory while preserving 16 supersymmetries.  This was  generalized in  \cite{Bena:2004jw} to  harmonic families of \nBPS{2} solutions.  The physics underpinning these solutions was that the M2 branes at infinity were becoming dielectrically polarized in the infra-red and giving rise to two sets of M5 branes  wrapping either one of the orthogonal $S^3$ factors in $S^3 \times S^3 \subset S^7$.  While these solutions were singular in both four and eleven dimensions, one could easily extract the physics of the solution once one obtained the uplift to M-theory.

Shortly after these solutions were constructed it was shown that these were special cases of much more general \nBPS{2} bubbling solutions, many of which could be arranged to be smooth in the infra-red limit \cite{Lin:2004nb}. Indeed, these ``LLM'' bubbling solutions are smooth, horizonless geometries supported by  cohomological fluxes in the infra-red and they helped guide the discovery of a similar geometric transition that led to ``bubbling black holes'' and black-hole microstate geometries in five dimensions.  The eleven-dimensional  LLM geometries have many possible infra-red limits that are governed by the sets of possible non-trivial cycles and magnetic $4$-form fluxes. These configurations were beautifully encoded in terms of a fermion droplet model that lives on a space transverse to all the branes. 

The initial motivation for this paper was an observation made in \cite{Bobev:2013yra} that the SU(3)$\times$\rm U(1)$\times$U(1)-invariant truncation of gauged $\Neql8$ supergravity has some similarities to the $\rm SO(4)$$\times$$\rm SO(4)$-invariant truncation that underlies the results in \cite{Pope:2003jp}.  Specifically, the supergravity potential in four dimensions has the same unexpected $\rm U(1)$ symmetry, which we will henceforth denote ${\rm U(1)}_\zeta$,  that relates very different M-theory solutions and very different holographic field theories.  The \suthuu-invariant truncation preserves, at most, $\Neql{2}$ supersymmetry  (the supersymmetries of the $\Neql{8}$ theory that are singlets under $\rm SU(3)$) and the bosonic sector of this truncation is exactly that of $\Neql{2}$ supergravity coupled to one vector multiplet. Indeed, we may view the underlying theory as precisely $\Neql{2}$ supergravity coupled to one vector multiplet with the  $\cR$-symmetry arising from one of the $\rm U(1)$ factors.

Unlike the solution studied in \cite{Pope:2003jp}, the superpotential considered here is only invariant under a $\ZZ_3$ subgroup of the unexpected ${\rm U(1)}_\zeta$ symmetry.  Thus the ${\rm U(1)}_\zeta$ symmetry will rotate solutions to the equations of motion into one another but will not generically preserve supersymmetry.  The initial goal was thus to study this rather more complicated class of solutions with an unexpected $\rm U(1)$ symmetry in the hope that it would admit generalizations as in  \cite{Bena:2004jw} and perhaps allow further generalization to smooth holographic geometries akin to those of LLM \cite{Lin:2004nb}.  Additionally, the goal was also to test the uplift formulae of \cite{deWit:1984nz,deWit:1986mz,deWit:1986iy,Nicolai:2011cy,deWit:2013ija,Godazgar:2013nma,Godazgar:2013pfa} and to lift the interesting class of Janus solutions given in \cite{Bobev:2013yra} to eleven dimensions. While these results will be detailed in a subsequent paper \cite{PTWnext}, the reason for this short note is to highlight some remarkable new flow solutions that emerge directly from  \suthuu-invariant truncation of gauged $\Neql8$ supergravity and lead to smooth infra-red limits that are dual to an ``almost conformal''  $(3+1)$-dimensional theory on intersecting M5-branes.  

The new solutions presented here have several  novel features.  First, they are  flows from a UV superconformal fixed point theory in $(2+1)$-dimensions to an infra-red theory that is an $S^1$ compactification (and non-trivial fibration) of a conformal fixed point theory in $(3+1)$-dimensions.  To be more specific, the UV fixed point is the usual theory on a stack of M2 branes and is thus dual to $AdS_4 \times S^7$.  Along the flow, the M2 branes polarize into M5 branes.  For the supersymmetric flows there is a choice of the parameter, $\zeta$,  in  ${\rm U(1)}_\zeta$ for which the M2 fluxes dissolve completely in the infra-red leaving only magnetic M5 fluxes.   These emergent M5 branes have $(3+1)$ directions in common:  the original M2 brane directions plus the $S^1$ of a non-trivial fibration over a six-dimensional, internal K\"ahler manifold, $\cB_6$.  The sets of M5 branes then further wrap two-dimensional sections of $\cB_6$.   At the UV fixed point, the manifold, $\cB_6$, is simply $\IC \IP_3$ and the fiber is simply the standard Hopf fibration but in the infra-red the K\"ahler potential and fiber connection are deformed away from that of $\IC \IP_3$ and $S^7$.   

Having obtained such a supersymmetric flow at a specific value of $\zeta$, we then use this solution and the ${\rm U(1)}_\zeta$ action to a find a family of non-supersymmetric flows with the same infra-red structure at generic values of $\zeta$. 

Remarkably, when the flow is arranged so that only M5's remain in the infra-red, the supergravity solution asymptotes to $\widehat {AdS_5} \times \cB_6$ where $\widehat {AdS_5}$ means that one takes the standard Poincar\'e metric on $AdS_5$ and periodically identifies one of the three spatial directions of the Poincar\'e slices and then fibers the resulting circle over $\cB_6$ using a connection induced from the K\"ahler potential of $\cB_6$.  Note that this circle still retains the same radial factor as the other space-time directions of the Poincar\'e slicing.  Thus the dual field theory is a $(3+1)$-dimensional conformal theory that lives on the common intersection of M5 branes but one of those directions is compactified, or periodically identified.  The non-trivial supergravity fibration presumably represents a  non-trivial  configuration, or twisting, of the fields {\it on} the M5 branes. 

The form of the $S^1$ fibration over $\cB_6$ that we encounter here is very reminiscent of the constructions in which M5 branes are wrapped on Riemann surfaces with a topological twist \cite{Maldacena:2000mw,Gaiotto:2009gz}.  The difference here though is that the twisted compactification of M5 branes emerges in the infra-red starting from M2 branes in the UV whereas, in the more standard construction, the M5 branes are  compactified and twisted in the UV and a new, lower-dimensional field theory emerges in the infra-red.  

At first glance, the holographic flows presented here might seem to challenge the usual wisdom of renormalization group flows:  the number of degrees of freedom should not increase as one flows to the infra-red and yet we appear to flow from a theory in $(2+1)$ dimensions to one in $(3+1)$ dimensions.  However, one should remember that we are looking at the infra-red limit of a strongly-coupled, large $N$ theory and its string theory dual and in such a context it is well-known that an extra dimension can emerge in the infra-red.  One of the more celebrated examples of this dates back to the early days of the ``second string revolution,'' where it was argued \cite{Witten:1995ex} that eleven-dimensional supergravity is an emergent low-energy, strong-coupling  phase of type IIA string theory.  In particular, the solitonic, charged black holes become massless in the strong coupling limit of the IIA theory and these new low-mass degrees of freedom may be interpreted as  the Kaluza-Klein modes of the emergent extra dimension.  Following on from this, in strongly-coupled $\cN=2$  Yang-Mills theory one can interpret the monopole and dyon spectrum as winding states of a little string in five or six dimensions \cite{Strominger:1995ac,Ganor:1996mu,Duff:1996cf,Seiberg:1996vs,Witten:1996qb,Klemm:1996bj,Ganor:1996nf}.   More recently,   there have been discussions of how extra dimensions can emerge from Coulomb branch flows for large $N$ theories.  (See, for example, \cite{HoyosBadajoz:2010td,Young:2014jma}.)  Thus, we expect that the emergence of the extra dimension in our flow should signal an interesting new strongly-coupled phase of the M2-brane theory in which families of charged, solitonic objects are becoming massless.  This may also have interesting implications for phases of the ABJM theory \cite{Aharony:2008ug} but one should note that the orbifold that leads to the ABJM theory will generically break the $\cN=2$ supersymmetry that underlies our supersymmetric flows.  On the other hand, as observed in \cite{Benna:2008zy}, such a relatively mild breaking of supersymmetry does not appear to modify other flows to infra-red fixed points.

 In Section \ref{Sect:4dsupergravity} we describe the details of the \suthuu-invariant truncation of gauged $\Neql8$ supergravity and find the BPS equations for these flows.  We also find some non-trivial integrals of the motion so that the entire family of flows can be solved implicitly via algebraic equations.  In Section \ref{Sect:11dsupergravity} we give the uplift of our flows to eleven-dimensional supergravity.   We then focus on one particular supersymmetric flow, with $\zeta$ fixed to $\pi/3$, because this has the most interesting infra-red behavior.  We show that the metric in the infra-red is that of  $\widehat {AdS_5} \times \cB_6$ and we show that only magnetic M5-brane fluxes survive in this limit.  We also note the apparent coincidence that our internal manifold $\cB_6$ is exactly the same as that of the non-trivial conformal fixed-point theory of \cite{Corrado:2001nv,Benna:2008zy}.  In Section \ref{Sect:11dsusy} we analyze the supersymmetry of our solutions and show how the dielectric polarization causes the supersymmetries to interpolate between those of M2 branes and those of M5 branes.    In Section  \ref{Sect:nonsusy} we generalize our interesting supersymmetric flow with $\zeta= \pi/3$ to non-supersymmetric flows at generic $\zeta$ but with the same interesting IR features.   Section \ref{Sect:Conclusions} contains our conclusions and a discussion of the possibilities for more general classes of flows like the one exhibited here.
  
\section{Gauged supergravity  in four dimensions}
\label{Sect:4dsupergravity}

\subsection{The action and BPS equations}
\label{ss:4dBPS}

The \suthuu-invariant sector of gauged $\Neql8$ supergravity contains one real scalar and one pseudo-scalar that combine naturally into a single complex scalar, $z$.\footnote{Our discussion here follows   \cite{Bobev:2013yra} where this and other truncations of gauged $\Neql8$ supergravity with the same scalar coset are presented in a unified framework.}  The scalar manifold is the    coset  $\rm SU(1, 1)/U(1)$ with the  K\"ahler form 
\begin{equation}
\mathcal{K} = - 3\, \log(1-z\bar{z})\,,
\label{Kahlerpot}
\end{equation}
and so the  scalar kinetic term is given by the canonical sigma-model expression:
\begin{equation}\label{Kform}
\mathcal{K}_{z\bar{z}} ~=~ \dfrac{3}{(1-z\bar{z})^2}\;.
\end{equation}

The holomorphic superpotential, $\cV(z)$, is a cubic: 
\begin{equation}
\mathcal{V} ~=~ \sqrt{2} (z^3 +1) \,,
\label{holV1}
\end{equation}
and the real superpotential, $W$, is defined by
\begin{equation}
W ~\equiv~   e^{\cK/2} \, |\cV| ~=~ \frac{\sqrt{2}\, |z^3 +1|}{(1-|z|^2)^{3/2}} \;.
\label{normW} 
\end{equation}
The potential, $\mathcal{P}(z,\bar{z})$, can be obtained from a holomorphic superpotential, $\cV(z)$, via:
\begin{equation}\label{Pform}
\begin{split}
\mathcal{P} = e^{\mathcal{K}}(\mathcal{K}^{z\bar{z}}\nabla_{z}\mathcal{V}\nabla_{\bar{z}}\overline{\mathcal{V}} -3 \mathcal{V}\overline{\mathcal{V}})\;,
\end{split}
\end{equation}
where the covariant derivatives are defined in the usual way:
\begin{equation}
\nabla_{z}\mathcal{V} = \partial_{z}\mathcal{V} + (\partial_{z}\mathcal{K})\mathcal{V}\;, \qquad \nabla_{\bar{z}}\overline{\mathcal{V}} = \partial_{\bar{z}}\overline{\mathcal{V}} + (\partial_{\bar{z}}\mathcal{K})\overline{\mathcal{V}}\;.
\end{equation}
Explicitly, one finds that the potential reduces to a  very simple result:
\begin{equation}
\mathcal{P}   ~=~ -\frac{6\,(1 + |z|^2)}{1-|z|^2}\;.
\label{PzSU3}
\end{equation}

Note that $\mathcal{P}$ does not depend upon the phase of $z$.  This is the unexpected ${\rm U(1)}_\zeta$ invariance.  It is unexpected  because, in eleven dimensions, the real part of $z$ (a four-dimensional scalar) comes from internal metric mode while  the imaginary part of $z$ (a four-dimensional pseudoscalar) comes from a mode of the tensor gauge field, $A^{(3)}$.  Thus rotating the phase of $z$ has an extremely non-trivial effect on the eleven-dimensional solution.  Also note that the superpotential, $W$, is only invariant under a $\ZZ_3$ subgroup of ${\rm U(1)}_\zeta$. 


To parametrize everything more explicitly, it is convenient to write the scalar in the following polar form:
\begin{equation}
z ~=~ \tanh \lambda\, e^{i\zeta}  \,.
\label{zparam}
\end{equation}
Then one has 
\begin{align}
W & ~=~  \sqrt{2}\, \sqrt{ \cosh^6 \lambda  +  \sinh^6 \lambda + 2\, \sinh^3 \lambda\,\cosh^3 \lambda\, \cos 3 \zeta}  \,,\label{Wreal}  \\[6 pt]
\mathcal{P} &  ~=~ -6\,  \cosh 2 \lambda \label{Preal}  \;. 
\end{align}
One can easily check that
\begin{equation}
\label{Simppot}
\cP~=~  \frac{1}{3}\,\bigg[ \left({\partial W \over\partial\lambda}\right)^2+ \frac{4}{\sinh^{2}(2\lambda)}\left({\partial W\over\partial\zeta}\right)^2\bigg] ~-~ 3 \, W^2 \,.
\end{equation}

As usual we take the four-dimensional space-time to consist of $(2+1)$-dimensional Poincar\'e  sections and a radial coordinate, $r$:
\begin{equation}
ds_4^2 ~=~ e^{2\,A(r)}\, \eta_{\mu \nu} dx^\mu dx^\nu  ~+~ dr^2\,.
\label{FlowMet}
\end{equation}
The effective particle action that encodes all field equations is then: 
\begin{equation}\label{su3lag}
\begin{split}
\cL~=~  & 3\, e^{3A}\left[(A')^2 ~-~ \dfrac{z'\bar{z}'}{(1-|z|^2)^{2}} ~+~ 2 g^2\,\Big(\dfrac{1+|z|^2}{1-|z|^2}\Big)
\right]    \\[6 pt]
~=~ &  e^{3A}\left[3 (A')^2  ~-~ 3\, (\lambda')^2  ~-~ {3\over 4}\sinh^2(2\lambda)(\zeta')^2~+~ 6g^2 \, \cosh(2\lambda)
\right]\,,
\end{split}
\end{equation}
where $g$ is the gauge coupling within the underlying gauged supergravity.


Note that the $\zeta$-independence of this action implies that there is a conserved momentum:
\begin{equation}
e^{3A}\, \sinh^2 2 \lambda  \,  \, \zeta'    ~=~ {\rm const.  }
\label{Nother1}
\end{equation}

The BPS equations are simply:
\begin{equation}
 A' ~=~  \pm g \,W \,, \qquad   \lambda'  ~=~   \mp {g \over 3}  \, {\partial W\over\partial \lambda} \,, \qquad 
\zeta' ~=~  \mp{ 4\,g \over 3 \, \sinh^2(2\lambda) }\,{\partial W\over\partial\zeta}  \label{4dBPS}\,.
\end{equation}
In the following we choose the top sign for which the UV fixed point is at $r\rightarrow \infty$.

\subsection{Integrating the flow}
\label{ss:Ioms}

\begin{figure}[t]
\centering
\includegraphics[width=8cm]{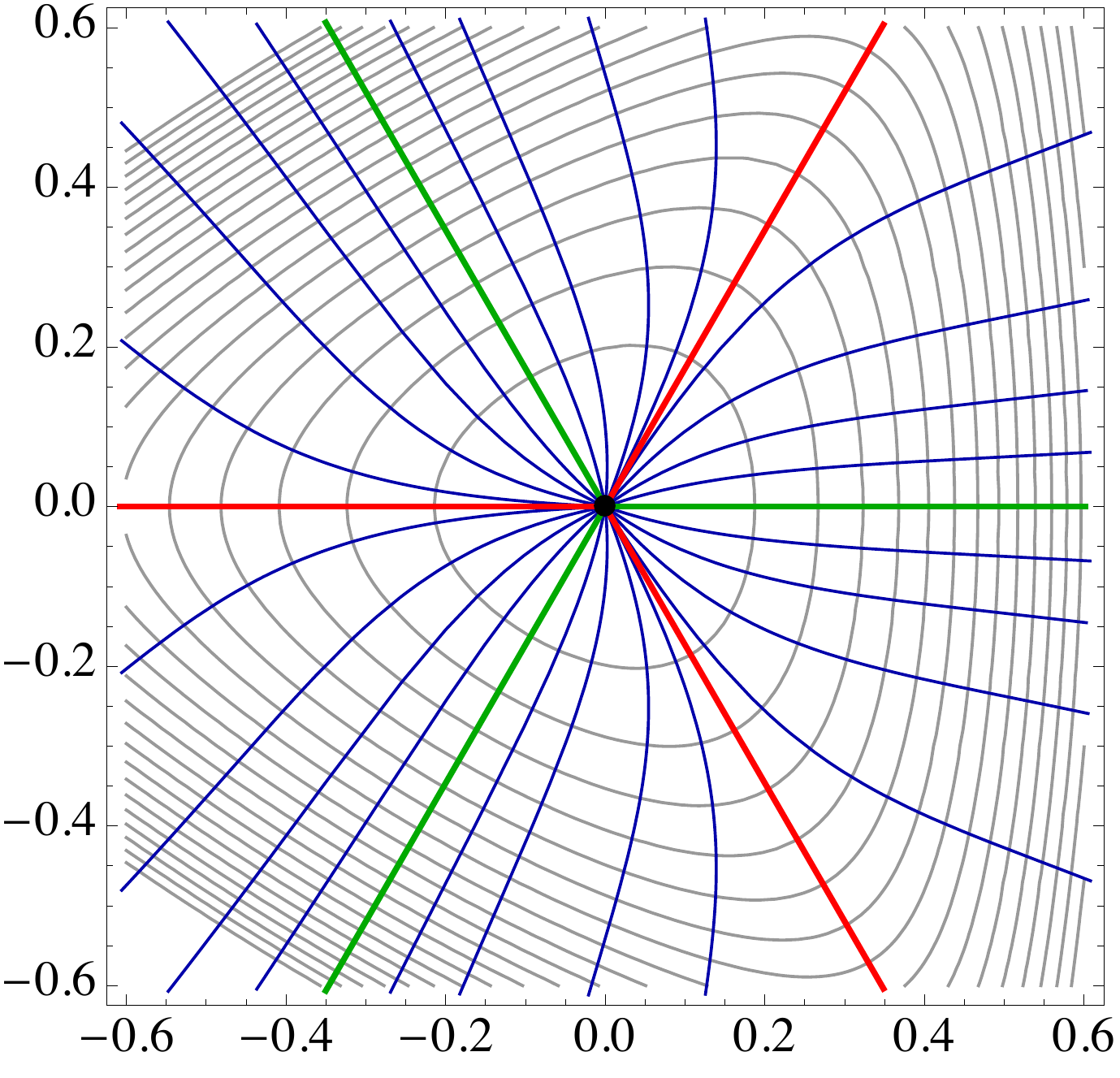}
\caption{RG-flow trajectories in the $(\lambda\cos\zeta, \lambda\sin\zeta)$-plane. The background contours are of the real superpotential $W(\alpha,\zeta)$. The ridge trajectories have constant $\zeta$ with $\cos 3\zeta=1$ (green) and $\cos 3\zeta=-1$ (red), respectively.
}
\label{Rgflows}
\end{figure}

Solutions to the BPS  equations \eqref{4dBPS} fall into two classes as illustrated in Figure \ref{Rgflows}: the generic flows with varying $\zeta$ and the ``ridge flows'' for special values of constant $\zeta$. Indeed, it follows immediately from  \eqref{4dBPS} that $\zeta$ is constant along a supersymmetric flow if and only if $\zeta =  n\pi/3$ for $n \in \ZZ$.  Moreover, one can also see that $\zeta' \to 0$, or $\zeta$ asymptotes to a constant value as $|\lambda| \to \infty$.

The generic solutions can be obtained algebraically using two integrals of  the motion:
\begin{equation} \label{Iom1}
\begin{split}
 \cI_1 & ~=~ e^{3A}\,{\partial W\over\partial\zeta}   ~=~ -e^{3A}\,\frac{3 \sqrt{2} \sinh ^3\alpha  \cosh ^3\alpha  \sin 3 \zeta }{\sqrt{2 \sinh ^3\alpha 
   \cosh ^3\alpha \cos 3 \zeta +\sinh ^6\alpha +\cosh ^6\alpha }} \,,
\end{split}
\end{equation}
and
\begin{equation}
\begin{split}
 \cI_2 ~=~ &    \frac{W^2}{(\cosh 2\lambda + \cos\zeta \, \sinh 2\lambda )^3  }   \,   \frac{\sin 3\, \zeta}{\sin^3 \zeta } \\[6 pt]
  ~=~ &  \frac{(4 \, \cos^2 \zeta -1)}{2 \,\sin^2 \zeta }\,\frac{(3+(\cosh 2\lambda -2 \cos\zeta \, \sinh 2\lambda )^2) }{(\cosh 2\lambda + \cos\zeta \, \sinh 2\lambda )^2  } \,. 
 \end{split} \label{Iom2}
\end{equation}
for the first order system \eqref{4dBPS}. 
Note that  $\cI_1$  is just the conserved momentum \eqref{Nother1} simplified using \eqref{4dBPS}.  The  second integral, $\cI_2$,   appears to be special and does not  correspond to any obvious symmetry of the Lagrangian \eqref{su3lag}. 

Given the two independent integrals of  the motion  for the BPS flow  and parametrizing everything in terms of the field $\lambda(r)$ rather than the ``unphysical coordinate,'' $r$, 
one can   use $\cI_2$ to obtain, implicitly, $\zeta(\lambda)$. In particular, $\cI_2$  relates the initial direction of the flow out of $\lambda =0$ to the asymptotic value as $\lambda \to \infty$. Using $\cI_1$ one can then determine $A(\lambda)$.  There are therefore no differential equations to be solved: Everything is determined implicitly from $\cI_2$ and  then explicitly from $\cI_1$.

\begin{figure}[t]
\centering
\includegraphics[width=7.5 cm]{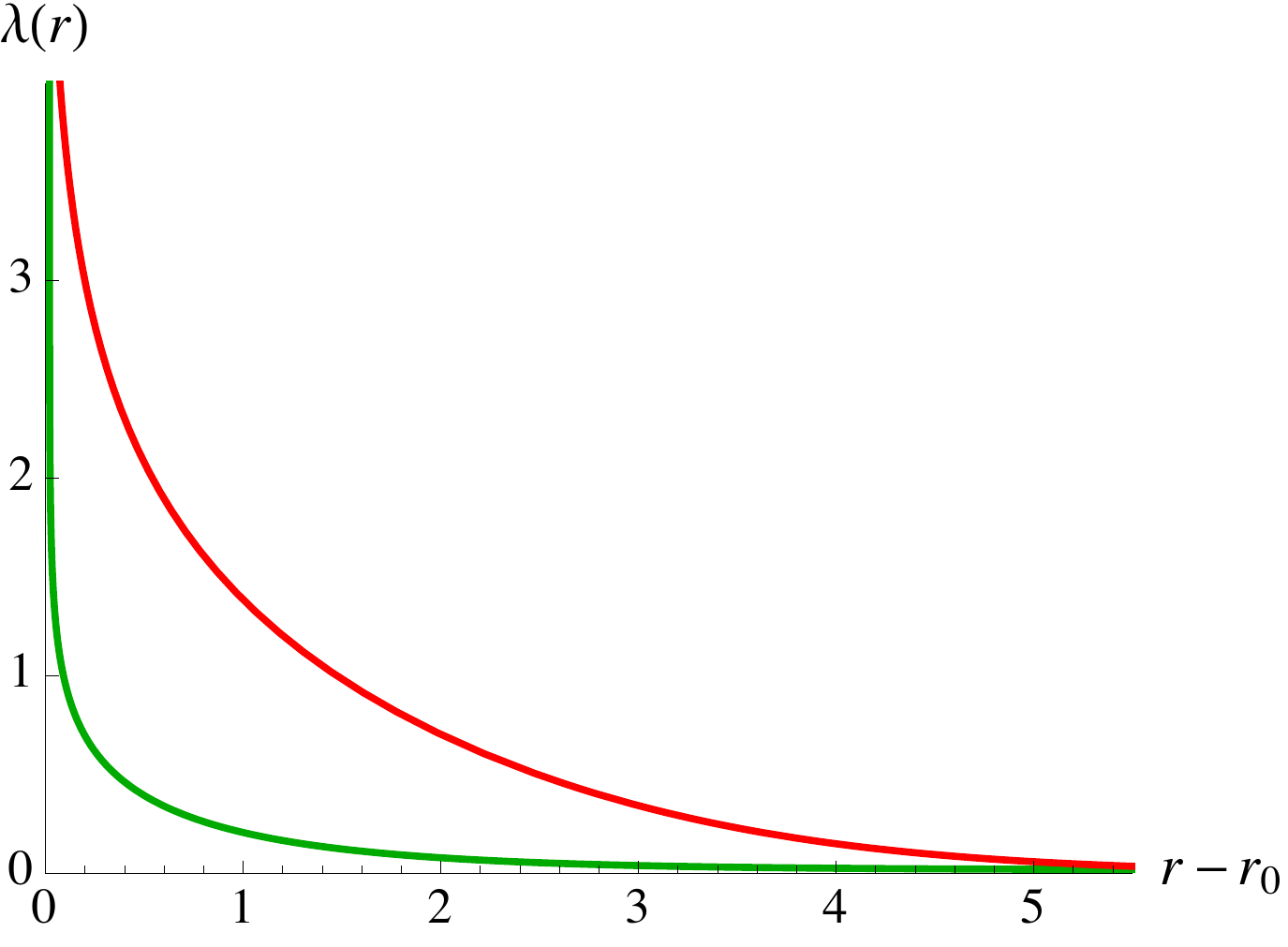}\qquad 
 \includegraphics[width=6 cm]{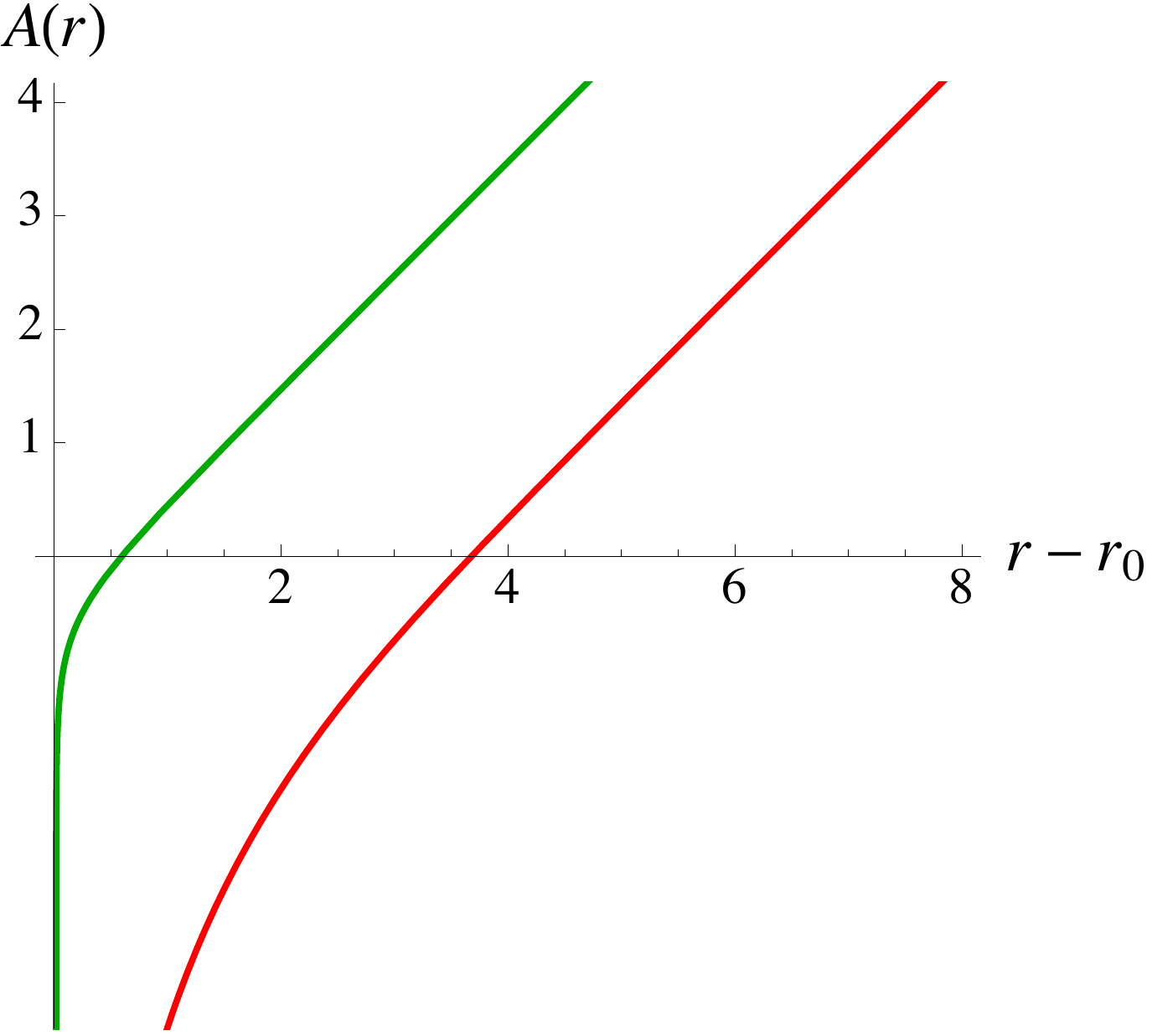}
\caption{Ridge flows for $\cos 3\zeta=-1$ (red) and $\cos 3\zeta=1$ (green) with  $A_0=0$.}
\label{ridgeflows}
\end{figure}

However, this  does not work for the ridge flows where  $\cI_1$ and $\cI_2$ either vanish or are singular and so do not directly yield a solution.  For such flows the first order system \eqref{4dBPS} reduces to the following two equations:
\begin{equation}\label{ridgeBPS}\begin{split}
A'& ~=~ \sqrt 2 \,g\,(\cosh^3\lambda\pm\sinh^3\lambda)\,,\\[6 pt]
 \lambda'& ~=~ -{g\over\sqrt 2} \sinh 2\lambda\,(\cosh\lambda\pm\sinh\lambda)\,,
\end{split}
\end{equation}
where the sign on the right-hand side is determined according to $\cos3\zeta =\pm 1$.
Using the UV boundary condition, $\lambda(r)\to 0$ as $r\rightarrow\infty$, we then find
\begin{equation}\label{sollambda}
\mathop{\rm arccoth}(e^{\lambda })\pm \arctan(e^{\lambda })\mp {\pi\over 2} ~=~ {g\over\sqrt 2}(r-r_0)\,,
\end{equation}
and
\begin{equation}\label{solA}
A(r) ~=~ A_0-\log(e^{4\lambda}-1)+
\begin{cases}
3 \lambda & \text{for $\cos 3\zeta=+1$}\\
\lambda &  \text{for $\cos 3\zeta=-1$}
\end{cases} 
\end{equation}
where $r_0$ and $A_0$ are integration constants.  The solutions \eqref{sollambda} and \eqref{solA} are regular for $r>r_0$, but become singular in the infra-red as $r\rightarrow r_0$. (See Figure~\ref{ridgeflows}.) It is this singularity that we want to explore further.

\subsection{Asymptotics}
\label{ss:asympflows}

We now consider the infra-red limit, $r\rightarrow r_0$,  in which $\lambda \to \infty$.  Using 
\begin{equation}\label{}
\mathop{\rm arccoth }x~\sim~{1\over x}+{1\over 3 x^3}+O\Big({1\over x^4}\Big)\,,\qquad 
\arctan x ~\sim~ {\pi\over 2}-{1\over x}+{1\over 3 x^3}+O\Big({1\over x^4}\Big)\,,\qquad x\rightarrow\infty\,,
\end{equation}
we obtain two different asymptotic expansions\footnote{Equivalently, one can simply solve \eqref{ridgeBPS} in the asymptotic region.}
 for the solutions
\eqref{sollambda} and \eqref{solA}:
\begin{align}
\cos 3\zeta = +1&:\quad \qquad  e^{ \lambda} ~\sim~    \frac{2 \sqrt{2}}{3g}  \,  \frac{1}{(r-r_0)}  \,,\qquad  A(\lambda)~\sim~-\lambda+A_0\,,\\[6 pt]
\cos 3\zeta = -1&:\quad \qquad e^{ \lambda} ~\sim~    \frac{2 \sqrt{2}}{g}  \,  \frac{1}{(r-r_0)}  \,,\qquad  A(\lambda)~\sim~-3\lambda+A_0\,.
\label{lamasymp2}
\end{align}
It turns out that the first expansion also applies to flows with generic $\zeta$. For the sake of brevity and interest, here we will focus only on  $\cos 3\zeta = -1$ and defer the more general discussion to  \cite{PTWnext}.


From the expansion \eqref{lamasymp2}, we find that  to leading order as $\lambda \to \infty$, the four-dimensional metric \eqref{FlowMet} becomes
\begin{equation}\label{asympmet2}
\begin{split}
ds_4^2 ~&\sim~ e^{-6\,\lambda}\, \eta_{\mu \nu} dx^\mu dx^\nu  ~+~  {8\over g^2}\,e^{-2\lambda}d\lambda^2 ~+~ \cO( e^{-10 \lambda} ) \\ & \qquad =~
{g^6\over 512} \, (r-r_0)^6 \, \eta_{\mu \nu} dx^\mu dx^\nu  ~+~  dr^2 ~+~ \cO(  (r-r_0)^{10})  \,. \end{split}
\end{equation}
%
The four-dimensional metric thus appears  to be very singular as $\lambda \to \infty$. For example, its Ricci scalar diverges as\footnote{For the generic flows, $R\sim -(3/4) g^2 e^{6\lambda}+O(e^{2\lambda})$, and the singularity is more severe.}
\begin{equation}\label{}
R~\sim~-{45\over 4}g^2\,e^{2\lambda}+O(e^{-2\lambda})\,.
\end{equation}
 As we will see, the eleven-dimensional solution is far better behaved.

\section{The solutions in M-theory}
\label{Sect:11dsupergravity}

The complete uplift of the foregoing RG-flows to solutions in M-theory can be obtained using the well-established uplift formula for the metric \cite{deWit:1984nz}  and the more recent  formulae for the flux \cite{deWit:2013ija,Godazgar:2013nma,Godazgar:2013pfa}. We have verified that the M-theory equations of motion are then indeed satisfied and that the supersymmetric flows in five dimensions lift to supersymmetric flows in eleven dimensions. This  works for all flows in Section~\ref{Sect:4dsupergravity}. The details of the calculation will be presented elsewhere  \cite{PTWnext}. In the following we summarize the results for general flows and then specialize to the ridge flows of interest.

\subsection{The metric uplift to eleven dimensions}
\label{ss:Metuplift}

\def\DX{X}

The  M-theory metric obtained by the uplift formula \cite{deWit:1984nz} is of the form
\begin{equation}\label{d11metr}
ds_{11}^2~=~ \Delta^{-1} ds_{4}+ds_7^2\,,
\end{equation}
where $ds_4^2$ is the metric \eqref{FlowMet}. By evaluating the internal metric, $ds_7^2$, and the warp factor, $\Delta$, and for a judicious choice of frames, we arrive  at the following result for the metric \eqref{d11metr}:
\begin{equation}
\begin{split}
e_i ~=~ &   \DX_+^\frac{1}{6} \, \Sigma^\frac{1}{3} \, e^{A(r)} \, dx_i \,, \quad i = 1,2,3  \,; \qquad   e_4  ~=~  \DX_+^\frac{1}{6} \, \Sigma^\frac{1}{3}  \, dr \,; \qquad   e_5  ~=~ a\, \DX_+^{-\frac{1}{3}} \, \Sigma^{\frac{1}{3} } \, d\chi \,;  \\[3 pt]
e_6 ~=~ &  a\, \DX_+^\frac{1}{6} \, \Sigma^{-\frac{1}{6}} \, \cos\chi \, d\theta \,; \qquad   e_7 ~=~    \frac{a}{2}\, \DX_+^\frac{1}{6} \, \Sigma^{-\frac{1}{6}} \, \cos\chi \,\sin \theta \,  \sigma_1 \,;     \\[3 pt]
e_8  ~=~   & \frac{a}{2}\, \DX_+^\frac{1}{6} \, \Sigma^{-\frac{1}{6}} \, \cos\chi \, \sin \theta \, \sigma_2 \,;   \qquad e_9 ~=~ \frac{a}{2} \, \DX_+^\frac{1}{6} \, \Sigma^{-\frac{1}{6}} \,  \cos\chi \,\sin \theta \,\cos \theta \, \sigma_3 \,;   \\[3 pt]
\qquad e_{10} ~=~ & a \, \DX_+^\frac{2}{3} \, \Sigma^{-\frac{2}{3}} \,  \sin \chi \,  \cos\chi \,\Big((d\psi + \coeff{1}{2}\, \sin^2 \theta\, \sigma_3) +  \coeff{(\DX_+ - \DX_-)}{\DX_+} \, d \phi  \Big) \,;  \\[3 pt]
e_{11} ~=~ &   a \, \DX_+^{-\frac{1}{3}} \, \Sigma^{-\frac{2}{3}} \,  \big(d \phi + \cos^2 \chi \,(d\psi + \coeff{1}{2}\, \sin^2 \theta\, \sigma_3)  \big) \,. \label{11dframs} 
\end{split}
\end{equation}
where $a$ is related to the four-dimensional gauge coupling constant \cite{deWit:1986iy},\begin{equation}\label{}
a={\sqrt 2\over g}\,,
\end{equation}
  and  the warp factors are given by:
\begin{equation}
 \DX_\pm(r)  ~\equiv~  \cosh 2\lambda \pm \cos \zeta  \,\sinh 2\lambda\,, \qquad  \Sigma (r,\chi) ~\equiv~   \DX_+\, \sin^2\chi +    \DX_- \,\cos^2\chi  \,.\label{defns1}
\end{equation}
It is also convenient to define the function
\begin{equation}
H_0(r,\chi)  ~\equiv~  \DX_+^\frac{1}{2} \, \Sigma \, e^{3\,A}    \,.\label{H0defn}
\end{equation}
Note that $H_0^{1/3}$ is the warp factor in front of the differentials $dx_i$.

The first four frames in (\ref{11dframs}) are simply the obvious frames for the four-dimensional metric (\ref{FlowMet}) multiplied by the warp factor $\Delta^{-1/2}=\DX_+^{1/6} \, \Sigma^{1/3}$.  The frames $e_6, \dots, e_9$ are frames for $\IC\IP_2$ multiplied by  $a \DX_+^{1/6} \, \Sigma^{-1/6} \cos \chi$ and the $\rm SU(3)$ symmetry acts transitively on this $\IC\IP_2$.  The one-form $(d\psi + \coeff{1}{2}\, \sin^2 \theta\, \sigma_3)$ is  the standard Hopf fiber over $\IC\IP_2$ that would extend it to $S^5$ and this one-form is also  $\rm SU(3)$  invariant.  The metric is invariant under $\psi$ and $\phi$ translations and these are the $\rm U(1)$ actions of the complete \suthuu\ invariance.   For $\lambda=0$, $e_5, \dots, e_{10}$ give the set of frames for $\IC\IP_3$ and $e_{11}$ is the Hopf fiber completing it to $S^7$.  To preserve the Poincar\'e invariance on the branes and the internal \suthuu\ symmetry, the warp factors and other fields can only have non-trivial dependence on  the internal coordinate, $\chi$, and the radial coordinate, $r$.

\subsection{The Maxwell potential}
\label{ss:Maxuplift}

The Maxwell potential, $A^{(3)}=A^{(3)}_{\rm st}+A^{(3)}_{\rm tr}$, necessarily has a term, $A^{(3)}_{\rm st}$,  proportional to the volume of the M2 branes. By the Poincar\'e and the \suthuu\ symmetry, it must be of the form
\begin{equation}\label{A3st}
A^{(3)}_{\rm st}~=~ h_0(r,\chi) \, e_1\wedge e_2 \wedge e_3\,,
\end{equation}
where $h_0(r,\chi)$ is a function that will be determined below. The potential, $A^{(3)}$,
 can also have components purely in the other eight directions transverse to the brane. We can choose a gauge in which the $dr$ components of  $A^{(3)}_{\rm tr}$ vanish, in other words $A^{(3)}_{\rm tr}$  can be arranged to have components only along the original $S^7$ directions.  Obviously $A^{(3)}_{\rm tr}$  must be \suthuu-invariant (up to gauge transformation).  This dramatically limits the possibilities but still leaves several functions of both $r$ and $\chi$ to be determined.  

Fortunately an uplift formula for $A^{(3)}_{\rm tr}$ in the same gauge has  been obtained in \cite{deWit:2013ija,Godazgar:2013nma,Godazgar:2013pfa}.  Applying it here and using the frames above, we arrive at a startlingly simple result:
\begin{equation}
A^{(3)}_{\rm tr}  ~=~   p(r)\, (e_6\wedge e_9 +e_7\wedge e_8  - e_5\wedge e_{10}  )\, \wedge e_{11} \,,\label{A3form}
\end{equation}
where 
\begin{equation}\label{pres1}
p(r) ~=~ {1\over 2}\sinh 2\lambda \,  \sin \zeta\,.
\end{equation}
It is somewhat surprising that the coefficient function, $p(r)$, is {\it only} a function of $r$: all the possible highly non-trivial $\chi$-dependence lies in the frames. 

Another very nice feature of the internal components of $A^{(3)}$ is that they have a symplectic structure that suggests a natural complex structure.  As we will see, this is not an accident.  In particular, $e_6\wedge e_9 +e_7\wedge e_8$ is proportional to the K\"ahler form on the $\IC\IP_2$.

The uplift of  the $A_{\rm st}^{(3)}$ part of the three-form potential is somewhat more involved since the existing uplift formulae are for the field strength, $F^{(4)}_{\rm st}=dA_{\rm st}^{(3)}$,  rather than for the potential itself. In particular, unlike in the uplifts of the metric and the transverse potential above, one must  use the BPS equations \eqref{4dBPS} to extract the function $h_0(r,\chi)$. 

The field strength, $F^{(4)}_{\rm st}$, can be obtained in a number of ways: One can use  the ``older'' uplift formula  in \cite{deWit:1986mz,deWit:1986iy,Nicolai:2011cy}, which gives the  part of $F^{(4)}_{\rm st}$ proportional to the four-dimensional volume  and then the Bianchi identity, $dF_{\rm st}^{(4)}=0$, to determine the remaining components. The result is
\begin{equation}\label{spflux}
F^{(4)}_{\rm st} ~=~ {g\over 3\sqrt 2}\,e^{3A}\,dx^1\wedge dx^2\wedge dx^3\wedge (U\,dr+V\,d\chi)\,,
\end{equation}
where
\begin{equation}\label{}
\begin{split}
U(r,\chi) & ~=~ -3(1-2\cos 2\chi)\sinh 2\lambda \cos\zeta-9\cosh 2\lambda\,,\\[6 pt]
V(r,\chi) & ~=~  {3\over 2g^2}\,\sin 2\chi\,(4\cos\zeta\,\lambda'-\sinh 4\lambda\sin\zeta\,\zeta')\,.
\end{split}
\end{equation}

Another method is to use the recent uplift formula for the dual six-form potential $A^{(6)}$ in \cite{Godazgar:2013pfa}. By dualizing its field strength, one arrives at the same result \eqref{spflux}! Finally, and computationally the simplest, is to use the BPS equations of M-theory from which, given  the  metric \eqref{d11metr} and the internal flux \eqref{A3form}, the remaining flux components are determined algebraically  \cite{PTWnext}.

One can integrate the field strength \eqref{spflux} to its potential \eqref{A3st} using the BPS equations \eqref{4dBPS}. This yields
\begin{align}\label{}
h_0(r,\chi) & ~=~  \frac{ \DX_+^{1/2}}{2\sqrt{2}\, W}\, \Big(  2\,\big(\cosh 2\lambda -\coeff{1}{2}\, \cos \zeta  \,\sinh 2\lambda\big)~-~  (2 - \cos 2\chi)\,\frac{\sinh^2 2\lambda \,  \sin^2 \zeta}{\Sigma} \Big) \,,\label{h0res}
\end{align}
and completes  the uplift of any supersymmetric \suthuu-invariant  RG-flow.

One should note that the uplift in \eqref{d11metr}, \eqref{A3form} and \eqref{spflux} yields a solution of the equations of motion in eleven dimensions for any domain wall type  solution (cf.\ \eqref{FlowMet}) of the equations of motion in four-dimensions that follow from the particle Lagrangian  \eqref{su3lag}. This means that we can still use these uplift formulae when we consider non-supersymmetric flows in Section~\ref{Sect:nonsusy}.    It is only in deriving the potential \eqref{A3st} and in the proof of supersymmetry in Section~\ref{Sect:11dsusy} that one uses the four-dimensional BPS equations \eqref{4dBPS}.

\subsection{IR asymptotics in eleven dimensions}
\label{ss:asymp11d}

Again we consider the limit in which $\lambda \to +\infty$ but now in eleven dimensions we find a very different picture. Here we focus on the supersymmetric solution  corresponding to the ridge flow  with $\zeta=\pi/3$. Note that reversing the sign of $\zeta$ amounts simply to  reversing the sign of the internal components of $A^{(3)}$, as is evident from \eqref{A3form}. For $\zeta=0, \pi$, the internal flux vanishes and the uplift given in  \eqref{d11metr} and  \eqref{spflux}  reduces to the known solution \cite{Cvetic:1999xx} dual to the Coulomb branch. 

Recall that $\zeta$ limits to a constant value at infinity.   
The various warp factors behave as follows as $\lambda \to +\infty$:
\begin{equation}
 \DX_\pm  ~\sim~ \coeff{1}{2} \,(1 \pm \cos \zeta) \, e^{2 \lambda}\,, 
 \qquad \Sigma  ~\sim~ \coeff{1}{2} \,  e^{2 \lambda} \, \widehat \Sigma \,, \qquad \widehat \Sigma \ ~\equiv~ (1 - \cos \zeta \, \cos 2\chi)
\label{11dasymp1}\,.
\end{equation}
%
%
Using \eqref{lamasymp2} we can change to the  more physical parametrization in terms of $\lambda$: 
\begin{equation}
d \lambda  ~\sim~ - \frac{1}{2 \,a} \, e^{ \lambda}\, dr  \,, 
 \qquad e^A  ~\sim~ R\,  e^{-3\, \lambda}   \label{11dasymp2}\,,
\end{equation}
where $R=e^{A_{0}}$ is an integration constant.  Then for $\zeta=\pi/3$ the leading terms in the metric are
\begin{equation}\label{11met1}
\begin{split}
ds_{11}^2  ~\sim~ 2^{-\frac{4}{3}}\, 3^\frac{1}{3}  \,\widehat \Sigma^\frac{2}{3}\, \bigg[\, &{  a^2}\, \frac{d\rho^2}{\rho^2}  ~+~  \rho^2 \, R^2\,(-dx_1^2 +dx_3^2 + dx_3^2)   \\[3 pt]  
& ~+~ \frac{64}{27  } \,\rho^2 \, {  a^2}\, \Big(d \phi + \frac{3\,\cos^2 \chi }{2\,\widehat \Sigma}\,(d\psi + \coeff{1}{2}\, \sin^2 \theta\, \sigma_3 +  \coeff{2}{3} \, d \phi) \Big)^2   \\[6 pt]
& ~+~ \frac{4}{3} \,a^2\, \Big( \, d \chi^2 ~+~ \frac{3 }{2\,\widehat \Sigma }\,\cos^2 \chi \, ds^2_{\IC\IP_2}  \\[3 pt]
& \qquad\qquad ~+~   \frac{9 }{4\, \widehat \Sigma^2 }\, \sin^2 \chi \,\cos^2 \chi \,\big(d\psi + \coeff{1}{2}\, \sin^2 \theta\, \sigma_3  +  \coeff{2}{3} \, d \phi\big)^2\,\Big) \,\bigg] \,, \end{split}
\end{equation}
where 
\begin{equation}
\rho ~\equiv~  e^{-2 \lambda} \,, \qquad   \widehat \Sigma \ ~\equiv~ (1 - \coeff{1}{2} \, \cos 2\chi)
 \label{redefs1}\,,
\end{equation}
and the subleading terms vanish as $\lambda\rightarrow\infty$.
Note that  this metric contains a factor corresponding to compact six-dimensional slice whose metric is conformal to 
\begin{equation}\label{6met1}
\begin{split}
\widehat{ds_6}{}^2 ~=~ & \frac{\widehat \Sigma }{ (1+\cos \zeta )}\,   d\chi^2 ~+~  \cos^2 \chi \, ds^2_{\IC\IP_2} \\[6 pt]
& ~+~   \frac{(1+\cos \zeta )  }{\widehat \Sigma }\, \sin^2 \chi \,\cos^2 \chi \,\Big(d\psi + \coeff{1}{2}\, \sin^2 \theta\, \sigma_3  +  \frac{2 \, \cos \zeta }{(1+\cos \zeta )} \, d \phi\Big)^2 \,, 
\end{split}
\end{equation}
for $\zeta = \pi/3$.  We have temporarily restored $\zeta$ because the appearance of this finite compactification manifold is a feature of all the flows.  This metric 
is K\"ahler with a K\"ahler form:
\begin{align}
\widehat J ~\equiv~  &- \sin \chi \,\cos \chi \, d\chi \wedge  \Big(d\psi + \coeff{1}{2}\, \sin^2 \theta\, \sigma_3  +  \frac{2 \, \cos \zeta }{(1+\cos \zeta )} \, d \phi\Big)  ~+~  \cos^2 \chi \, J_{\IC\IP_2} \\[6 pt]
~=~  & d\, \Big[\coeff{1}{2}\, \cos^2 \chi \, \Big(d\psi + \coeff{1}{2}\, \sin^2 \theta\, \sigma_3  +  \frac{2 \, \cos \zeta }{(1+\cos \zeta )} \, d \phi\Big) \Big]
 \,, \label{Kform1}
\end{align}
where $J_{\IC\IP_2}$ is the K\"ahler form on $\IC\IP_2$.\footnote{This K\"ahler metric arises here by the same mechanism as in the general construction in \cite{Page:1985bq} (also see \cite{Gauntlett:2004hh}).}

Many of the components of $F$ arising from (\ref{A3form}) and  (\ref{h0res}) vanish in the infra-red and we find that this limiting Maxwell field is simply given by 
\begin{equation}
F^{(4)}_{{\rm IR}} ~=~   d A_{\rm IR}^{(3)}  \,, \qquad  A_{\rm IR}^{(3)}  ~=~ \frac{\sqrt{3} \,a^3}{4\, \widehat \Sigma} \,  \cos^4 \chi \ J_{\IC\IP_2}\wedge (d\psi + \coeff{1}{2}\, \sin^2 \theta\, \sigma_3 +  \coeff{2}{3} \, d \phi)
 \label{IRFsimp2}\,.
\end{equation}
Note, in particular, that the space-time components parametrized by $h_0$ vanish in this limit and so $F^{(4)}_{{\rm IR}}$ is purely magnetic and lives entirely on the conformally K\"ahler six-manifold.  The infra-red limit is thus sourced entirely by M5 branes.



The first and rather remarkable surprise is that the warp factor, $H_0^{1/3}$, defined by  (\ref{H0defn}) and lying in front of frames parallel to the M2-branes  (\ref{11dframs}),  is not singular in the infra-red for $\zeta \ne 0, \pi$.  For  $\zeta = 0, \pi$,   $H_0$ is expected to be singular because such a flow has no internal fluxes and $H_0$ is a harmonic function describing M2 branes spreading on a Coulomb branch.  However  there is no singularity for generic $\zeta$ and, for $\zeta =\pm \pi/3$,  (\ref{11met1}) shows that this warp factor actually vanishes in the IR (as is required for an AdS geometry).  Thus there are no singular sources of M2 branes in the infra-red.

The second surprise is that all the M2-brane fluxes, in fact, vanish in the infra-red and all that remains is a very simple non-singular magnetic  (M5-brane) flux (\ref{IRFsimp2}) on a  {\it finite-size}, conformally  K\"ahler, six-dimensional manifold, $\cB_6$.   One can also easily verify that as $\chi \to \pi/2$ this manifold is smooth.      The limiting metric (\ref{11met1}) is almost like that of $AdS_5 \times  \cB_6$:   The five dimensional manifold that we label as $\widehat{AdS_5}$ is $AdS_5$ in  Poincar\'e form with one spatial direction compactified and fibered over $\cB_6$.   Holographically it suggests that the infra-red phase is almost a CFT except that one spatial direction has been ``put in a periodic box'' of some fixed scale and that some interactions have been turned on so that this direction becomes non-trivially fibered.  Thus the infra-red phase is almost a CFT fixed point.

There is evidently a geometric transition in which the M2 branes  dissolve into smooth M5-brane fluxes leaving a finite size ``bubble'' in the form of a six-dimensional K\"ahler manifold.    This is very reminiscent of the kind of transition one finds in microstate geometries \cite{Bena:2005va,Berglund:2005vb,Bena:2007kg}.    

As we will see below, the supersymmetry analysis shows that the infra-red limit contains three species of M5 branes.  These branes all share  the original M2-brane directions {\it and} the fiber direction defined by $e_{11}$.  The three different species are then distinguished by how the two remaining directions wrap the three sets of complex directions in  $\cB_6$.  Thus the infra-red physics is entirely dominated by families of M5 branes with $(3+1)$ common directions, one of which is compactified.  This is why there are now three spatial directions on the same footing when it comes to the radial scaling and this is what gives rise to the $\widehat{AdS_5}$ factor.

Interestingly enough, the metric (\ref{11met1}) also has warp factors arranged so that if one performs a dimensional reduction to {\it five} dimensions on $\cB_6$ then the resulting five-dimensional metric is exactly that of  $\widehat{AdS_5}$  without any factors of $\widehat \Sigma$.  Thus the infra-red supergravity limit may really be thought of as a truly five-dimensional theory dual to a compactified and twisted, conformal $(3+1)$-dimensional field theory on intersecting M5 branes.

\subsection{The infra-red fixed point of the CPW flow}
\label{ss:CPW}

There is something of a coincidence that is worth noting at this juncture.  There is another apparently unrelated but well-known holographic flow to a conformal fixed point in  (2+1)~dimensions \cite{Ahn:2000mf,Corrado:2001nv}.  This involves turning on  rather different pseudoscalars in four dimensions corresponding to different eleven-dimensional fluxes, and  flowing to a solution that is
$AdS_4 \times \cM_7$.   The infra-red limit of this flow has an eleven-dimensional metric \cite{Corrado:2001nv}:
\begin{align}
ds_{11}^2  ~\sim~ \frac{1}{3} \,a^2\, \widehat \Sigma^\frac{2}{3}\, \bigg[\, \frac{d\rho^2}{\rho^2} & ~+~ \frac{3\,\sqrt{3}}{ a^2}\, \rho^2 \, (-dx_1^2 +dx_3^2 + dx_3^2) \nonumber \\  
& ~+~  \bigg(d \phi + \frac{3\,\cos^2 \chi }{2\,\widehat \Sigma}\,(d\psi + \coeff{1}{2}\, \sin^2 \theta\, \sigma_3 +  \coeff{2}{3} \, d \phi) \bigg)^2 \nonumber \\
& ~+~2 \, \Big( \, d \chi^2 ~+~ \frac{3 }{2\,\widehat \Sigma }\,\cos^2 \chi \, ds^2_{\IC\IP_2} \nonumber\\
& \qquad\qquad ~+~   \frac{9 }{4\, \widehat \Sigma^2 }\, \sin^2 \chi \,\cos^2 \chi \,\big(d\psi + \coeff{1}{2}\, \sin^2 \theta\, \sigma_3  +  \coeff{2}{3} \, d \phi\big)^2\,\Big) \,\bigg] \,, \label{11met2}
\end{align}
where $\widehat \Sigma$ is exactly as in (\ref{11dasymp1}) with $\zeta =  \pi/3$.  Note that the overall warp factor, compact fiber and base metric on $\cB_6$ are all {\it exactly} the same as those of our flow (\ref{11met1}).

This seems a remarkable coincidence and perhaps reflects some similar core physical relationship between the infra-red fixed point theory of \cite{Corrado:2001nv,Benna:2008zy} and the infra-red theory dual to the solution obtained here. On the other hand it is possible that this coincidence merely reflects the restricted class of conformally K\"ahler metrics that can be accessed within the consistent truncation to $\Neql8$ supergravity in four dimensions.   It would also be interesting to understand how the flows considered in this paper relate to the observations in \cite{HoyosBadajoz:2010td} about higher-dimensional Coulomb branch flows related to the CPW flow.

\section{The supersymmetry in eleven dimensions}
\label{Sect:11dsusy}

Since we are dealing with an $\Neql2$ supersymmetric solution arising through brane polarization, it is natural to expect that the supersymmetries of our solution to be defined  most naturally through projectors that reflect the underlying almost-complex structure and a dielectric projector much like those encountered in   \cite{Gowdigere:2003jf,Pilch:2004yg,Nemeschansky:2004yh,Pope:2003jp}.

\subsection{Defining the supersymmetries through projectors}
\label{ss:projectors}

To define the four residual supersymmetries along our flow we will require that the eleven-dimensional supersymmetry, $\epsilon$, obeys three (commuting) projection conditions:
\begin{equation}
\Pi_0 \, \epsilon ~=~\Pi_1 \, \epsilon ~=~ \Pi_2 \, \epsilon ~=~ 0 \,.
 \label{projconds}
\end{equation}
We also require that $\epsilon$ be independent of all coordinates except, possibly, $r$, $\chi$, $\phi$ and $\psi$.  As we will see below, the dependence on $\phi$ and $\psi$ is easily determined through the actions of the U(1)  symmetries.

The first, and most obvious  projector arises from the fact that the Killing spinor must be a singlet under the holonomy group, $\rm SU(3)$, of $\IC \IP_2$. The simplest way to implement this is to take:
\begin{equation}
\Pi_1 ~\equiv~ \frac{1}{2}(1+\Gamma^{6789}) 
 \label{Pi1}\,,
\end{equation}
where $e^6,\dots, e^9$ are the frames along $\IC \IP_2$.  

The second projector comes from assuming that another pair of frames extend the complex structure of  $\IC \IP_2$  to some form of almost-complex structure.  {\it A priori} one does not know exactly how this extra complex pair of frames emerges from $e^4,e^5,e^{10},e^{11}$ and so one has to allow some frame rotations to set up the projectors.  To do this, we define:
\begin{align}
\Gamma^{\widehat{4}}&~\equiv~\cos\alpha\, \Gamma^{4}+\sin\alpha\, \Gamma^{5}\,, \qquad
\Gamma^{\widehat{5}}~\equiv~\cos\alpha\, \Gamma^{5}-\sin\alpha\, \Gamma^{4}\,, \\[6 pt]
\Gamma^{\widehat{10}}&~\equiv~\cos\omega\, \Gamma^{10}+\sin\omega\, \Gamma^{11}\,, \qquad
\Gamma^{\widehat{11}} ~\equiv~\cos\omega\, \Gamma^{11}-\sin\omega\, \Gamma^{10}\,, 
\end{align}
where $\alpha =\alpha(r,\chi)$ and $\omega =\omega(r,\chi)$ are, as yet, undetermined functions.

We then take
\begin{equation}
\Pi_2 ~\equiv~  \frac{1}{2}(1-\Gamma^{\widehat{5}78\widehat{10}})
 \label{Pi2}\, .
\end{equation}

Finally there is the dielectric projector.  It is required to commute with $\Pi_1$ and $\Pi_2$ and be a dielectric deformation of the standard M2-brane projector.  This constrains the form significantly, but does not uniquely fix it.  There are essentially two different classes of such projectors: the ones that are variations on those used in  \cite{Gowdigere:2003jf,Pilch:2004yg,Nemeschansky:2004yh,Pope:2003jp} and something that is equivalent to taking:
\begin{equation}
\Pi_0 ~\equiv~  \frac{1}{2}(1+\cos\beta\,\Gamma^{123}+\sin\beta\, \Gamma^{\widehat{4}}) 
\label{Pi0}\,,
\end{equation}
where  $\beta =\beta(r,\chi)$ is some, as yet, undetermined function.  It turns out that this is the correct choice.

Using the identity $\Gamma^{1\ldots 11}=1$ and the projectors (\ref{Pi1}) and (\ref{Pi2}), we can replace $\Gamma^{\widehat{4}}$ in (\ref{Pi0}) by any one of the following
\begin{equation}
\Gamma^{123 6 9\,\widehat{11}} \,, \qquad \Gamma^{123  7 8\,\widehat{11}} \,, \qquad  \Gamma^{123 \widehat{5} \widehat{10}\,\widehat{11}}  \,.
\label{Pi0parts}
\end{equation}
These projector components make the interpretation of our projection conditions  more apparent: $\Pi_0$ is a dielectric projector reflecting the polarization of the original M2 branes into  three sets of M5 branes and  the angle $\beta$ reflects the extent of the polarization.  As mentioned earlier, these sets of M5 branes have four common frame directions, $1,2,3$ and $\widehat{11}$, and also wrap one of each pair of complex frames $(\widehat{5}, \widehat{10})$, $(6,9)$ or $(7,8)$.  Thus these M5 branes intersect on a common $(3+1)$-dimensional space.

When only the M5 branes survive in the infra-red, then the common intersection space of the M5 branes, defined by $e^1,e^2,e^3$ and $e^{\widehat{11}}$, must all be on the same footing within the solution. It is the field theory on this common intersection that leads to the  $(3+1)$-dimensional infra-red fixed point theory whose dual is the ``Poincar\'e''  $\widehat{AdS_5}$  factor in (\ref{11met1}). 

\subsection{The supersymmetry along the flow}
\label{ss:11dflowsusies}

We find that the supersymmetries defined above satisfy the eleven-dimensional BPS conditions provided that we take: 
\begin{equation}\label{}
\begin{split}
\cos \alpha  & ~=~ \frac{  (2\cos 2\chi -1 ) \DX_+^{1/ 2} } {\Omega^{1/2}}\,,\qquad 
\sin\alpha ~=~ -\frac{ \sqrt 2 \sin 2\chi \,W} {\Omega^{1/2}}\,,\\[6 pt]
\cos\omega & ~=~  -\frac{  (2-\cos 2\chi )\DX_+^{1/2} }{\Omega^{1/2}}\,,\qquad 
\sin\omega  ~=~ \frac{  \sin 2\chi (\DX_+-3\DX_-)\DX_+^{1/2} }{2\,\Omega^{1/2}}\,,
\end{split}
\end{equation}
and 
\begin{equation}\label{}
\begin{split}
\cos \beta  & ~=~- \frac{ {\DX_+^{1/2}}}{2\sqrt{2}\,\Sigma\, W }(\DX_+^2\sin^2\chi +(\DX_+\DX_- - 2) (\cos 2\chi -2)+3\DX_-^2\cos^2\chi) \,,\\[6 pt]
\sin \beta  & ~=~ \frac{\sinh 2\lambda \sin \zeta}{\sqrt{2}\,\Sigma\, W }\, \Omega^{1\over 2} \,,
\end{split}
\end{equation}
where
\begin{equation}\label{}
\Omega~=~ \DX_+(2\cos 2\chi -1 )^2 +2 \sin^2 2\chi\, W^2 \,.
\end{equation}
It easy to show that $\cos\beta$ is related to the time component of the flux by simple relation: 
\begin{equation}
h_0(r,\chi) ~=~ - \frac{1}{2}\, \cos\beta \,.
\label{h0cosbeta}
\end{equation}

With these angles we find that the fluxes and metric defined in Section~\ref{Sect:11dsupergravity} satisfy the eleven-dimensional BPS conditions and equations of motion provided that one uses the four-dimensional BPS equations, (\ref{4dBPS}).

\subsection{The infra-red supersymmetries }
\label{ss:IRsusies}

Again we consider the limit $\lambda \to \infty$ for $\zeta = \pi/3$.    We find in this limit: 
\begin{equation}
h_0(r,\chi) ~=~  -\frac{1}{2}\, \cos\beta  ~=~0 \,, \qquad \omega~=~ 0\,,
\label{h0van}
\end{equation}
and 
\begin{equation}
\cos \alpha ~=~  \frac{  2\cos 2\chi -1 } {2-\cos 2\chi} \,, \qquad \sin \alpha ~=~ -  \frac{ \sqrt{3} \sin 2\chi } {2-\cos 2\chi}\,.
\end{equation}

Thus we see that $\cos\beta$ vanishes precisely because it is proportional to the M2-flux function, $h_0(r,\chi)$, and this reflects the fact that there are only M5 fluxes in the infra-red.   The projector (\ref{Pi0}) becomes  $\Pi_0=\frac{1}{2}(1+\Gamma^{\widehat{4}})$ and, as we noted above, we can replace $\Gamma^{\widehat{4}}$ by any of (\ref{Pi0parts}) and this is consistent with the infra-red limit only consisting of M5 branes and their fluxes as in (\ref{IRFsimp2}). The fact that $\omega =0$ means that $\Gamma^{\widehat{10}}=\Gamma^{10}$ and $\Gamma^{\widehat{11}}=\Gamma^{11}$ and hence  the common internal M5 direction is indeed defined by $e^{\widehat{11}}=e^{11}$, which is consistent with the asymptotic form of the metric (\ref{11met1}).

Thus the structure of the supersymmetries in the infra-red is completely consistent with the interpretation coming from the flux and metric analysis.

\section{Non-supersymmetric flows}
\label{Sect:nonsusy}

We now return to the original reason that piqued our interest in this family of flows: The supergravity potential is completely independent of $\zeta$ while  the uplift to eleven dimensions and the physics have very non-trivial dependence on $\zeta$.  This suggests that might be a parallel  with the \nBPS{2} story \cite{Pope:2003jp,Bena:2004jw,Lin:2004nb} leading to  brane polarization and smooth geometries in the infra-red limit of the flow.  What makes the difference here is that, unlike  the example in \cite{Pope:2003jp}, the superpotential depends upon $\zeta$ and so supersymmetric flows are sensitive to the phase of the pseudoscalar.  However, we can still find families of interesting flows with $\zeta$ arbitrary and fixed: they are simply not supersymmetric for general $\zeta$.   

One should recall that any solution to the BPS equations (\ref{4dBPS}) automatically solves the {\it equations of motion} derived from  (\ref{su3lag}).  If one sets $\zeta = \pm \pi/3$ and, as before, takes the top choice of sign in  (\ref{4dBPS}), then the BPS equations explicitly yield (as in \eqref{ridgeBPS}):
\begin{equation}
 A' ~=~   \sqrt{2}\, g \,  (\cosh^3 \lambda  -  \sinh^3 \lambda)  \,, \qquad   \lambda'  ~=~   - \sqrt{2}\, g  \, e^{-\lambda} \cosh  \lambda \sinh \lambda   \,, \qquad 
\zeta' ~=~ 0   \label{spec4dBPS}\,.
\end{equation}
It follows that the solutions to the equations (\ref{spec4dBPS}) also solve the equations of motion for {\it any fixed value of $\zeta$}.  

Flows described by  (\ref{spec4dBPS}) are  non-supersymmetric unless $\zeta = \pm \pi/3$.  Moreover, apart from the lack of supersymmetry,  our other inferences and conclusions largely remain valid.   First, as $\lambda \to +\infty$  we have
\begin{equation}
e^{ \lambda} ~\sim~    \frac{2 \sqrt{2}}{g}  \,  \frac{1}{(r-r_0)}  \,,\qquad  A(\lambda)~\sim~-3\lambda+A_0\,,
\label{lamasymp2a}
\end{equation}
and  
\begin{equation}
X_\pm   ~\sim~ \coeff{1}{2} \,(1 \pm  \cos \zeta) \, e^{2 \lambda}\,, \qquad  
 \qquad \Sigma  ~\sim~ \coeff{1}{2} \,  e^{2 \lambda} \, \widehat \Sigma \,, \qquad \widehat \Sigma \ ~\equiv~ (1 - \cos \zeta \, \cos 2\chi)
\label{11dasymp1a}\,,
\end{equation}

The uplift formulae are almost all {\it algebraic} in the scalar fields.  This means that the eleven-dimensional metric coming from (\ref{11dframs}) is asymptotic to 
\begin{equation}\label{11met1a}
\begin{split}
ds_{11}^2  ~\sim~\frac{1}{2} \,& (1+\cos\zeta)^\frac{1}{3}  \,\widehat \Sigma^\frac{2}{3}\, \bigg[\, {  a^2}\, \frac{d\rho^2}{\rho^2}  ~+~  \rho^2 \, R^2\,(-dx_1^2 +dx_3^2 + dx_3^2)   \\[3 pt]  
&  +~ \frac{8}{(1+\cos\zeta)^3 } \,\rho^2 \, {  a^2}\, \Big(d \phi + \frac{(1+\cos\zeta) }{\widehat \Sigma}\,\cos^2 \chi \,\Big(d\psi + \coeff{1}{2}\, \sin^2 \theta\, \sigma_3 +   \frac{2\cos\zeta}{(1+\cos\zeta) }\,  d \phi\Big) \Big)^2   \\[6 pt]
& +~ \frac{2}{(1+\cos\zeta)} \,a^2\, \bigg( \, d \chi^2 ~+~ \frac{(1+\cos\zeta) }{\widehat \Sigma }\,\cos^2 \chi \, ds^2_{\IC\IP_2}  \\[3 pt]
&  +~   \frac{(1+\cos\zeta)^2}{ \widehat \Sigma^2 }\, \sin^2 \chi \,\cos^2 \chi \,\Big(d\psi + \coeff{1}{2}\, \sin^2 \theta\, \sigma_3  +   \frac{2\cos\zeta}{(1+\cos\zeta) }\,    d \phi \Big)^2 \, \bigg)\,\bigg] \,. \end{split}
\end{equation}
This is, once again, of the form $\widehat{AdS_5} \times \cB_6$ where the metric on $\cB_6$ is conformally K\"ahler.  Indeed, note that the metric on the $\cB_6$ sections is given precisely by a conformal multiple of (\ref{6met1}) with K\"ahler form (\ref{Kform1}).  This time, however, $\zeta$ is arbitrary.  

The internal fluxes, (\ref{A3form}), only depend upon the  function, $p$, in (\ref{pres1}) and so the large-$\lambda$ asymptotics for generic fixed $\zeta$ is much as before.  The only part of the uplift formula that depends upon derivatives of fields are the space-time components of the Maxwell fields, given by (\ref{spflux}).  It is, however, easy to integrate (\ref{spflux}) based upon (\ref{spec4dBPS}) for fixed $\zeta$ and we find
\begin{equation}\label{h0nonsusy}
h_0  ~=~-\frac{3}{\sqrt{2}\, g} \,X_+^{-\frac{1}{2}}\, e^{-\lambda} \, \Big(1+ \frac{(1 + \cos \zeta)}{2\,\Sigma}  \sinh 2\lambda  \Big)\,.
\end{equation}
From this it follows that the space-time components of the Maxwell field still die off at infinity as $e^{-2 \lambda}$ and so there are also no M2 branes in the core of these non-supersymmetric flows.

We conclude that all the essential physical features of the supersymmetric flows with $\zeta =  \pi/ 3$ remain true for the non-supersymmetric flows defined by (\ref{spec4dBPS}) for general $\zeta$.  The M2 branes at infinity dissolve into a cloud of M5 branes, leaving no M2 branes.  While we do not have the guidance from the supersymmetry projectors, the asymptotic form of the metric and fluxes suggest that the distributions of the M5 branes is also similar and, in particular, the M5 branes all share $(e^1,e^2,e^3, e^{11})$ as common directions leading to an ``almost conformal'' $(3+1)$-dimensional field theory in the IR.

\section{Conclusions}
\label{Sect:Conclusions}

One of the original motivations for studying the\suthuu-invariant flow was the possibility of parallels with the \nBPS{2} story \cite{Pope:2003jp,Bena:2004jw,Lin:2004nb} leading to  brane polarization and smooth geometries in the infra-red limit of the flow.   It is also interesting to note that the potential, $\cP$, given in (\ref{Preal}) is always bounded above and so satisfies the ``good'' condition of \cite{Gubser:2000nd} for apparently singular ``flows to Hades'' in the infra-red.   Thus one has a better expectation of finding something more physical in such an infra-red limit, as indeed we do.

We have largely focussed on the most interesting example of a supersymmetric flow with $\zeta = \pi/3$ because it exhibits a remarkable infra-red limit in which only M5-brane fluxes survive and whose dual field theory is almost conformal in $3+1$ dimensions.  The infra-red geometry, $\widehat {AdS_5} \times \cB_6$, is smooth up to possible orbifold singularities at fixed points of the $\rm U(1)$ orbits.  In the dielectric polarizations  of the \nBPS{2} geometries \cite{Pope:2003jp,Bena:2004jw}, it required a lot more work to find the smooth bubbled geometries  \cite{Lin:2004nb}, however, here we have found examples directly within  gauged $\Neql8$ supergravity and this is somewhat remarkable.

We have also used the special ($\zeta = \pi/3$) class of supersymmetric flows to construct non-supersymmetric flows that exhibit exactly the same remarkable infra-red structure but at general values of $\zeta$.  The existence of such non-supersymmetric flows is a simple consequence of the fact that the scalar potential, $\cP$, is independent of $\zeta$ while the superpotential depends upon $\zeta$.  The non-supersymmetric flows coincide with the supersymmetric ones when  $\zeta = \pm \pi/3$.

The other supersymmetric flows ($\zeta \ne \pi/3$) are also interesting and will be discussed further in \cite{PTWnext}, here we will simply comment that the infra-red limit is also  smooth (up to orbifolds)  but rather  different from the one encountered here. For generic $\zeta$, there is still some M2 flux surviving in the infra-red limit, the compactification manifold, $\cB_6$, is still conformally K\"ahler and the scale factors for this part of the metric  {\it and}  the metric along the M2 branes limits to a finite value.  The radial coordinate and the remaining $\rm U(1)$ fiber conspire to make an orbifold of $\IR^2$ fibered over $\cB_6$.  There is no hint of conformal behavior in the infra-red and the  $\rm U(1)$ fiber scales to zero size while the M2-brane part of the metric retains a finite scale, which means that the infra-red limit is still intrinsically a $(2+1)$-dimensional field theory on the M2 branes.

For holography, the class of flows considered here are extremely interesting.  They start from the $(2+1)$-dimensional conformal field theory of M2 branes and ``flow up in dimensions'' to a field theory on the common $(3+1)$-dimensional intersections of M5 branes.  Moreover, the theory on the M5 branes is almost conformal:  the $AdS_5$ has one spatial direction compactified and fibered.  The compactification is relatively straightforward but presumably the non-trivial fibration corresponds to adding background fields and twisting the theory on the M5 branes.  As we discussed in the introduction, we expect that this signals a new strongly-coupled phase of the holographic field theory on the M2 branes in which families of charged solitonic objects are becoming massless.   It would be extremely interesting to understand this in more detail for the M2-brane theory and for its orbifolds that lead to the ABJM theory \cite{Aharony:2008ug}.

From the pure supergravity side, this work also provides the first and a highly non-trivial application of the uplift formulae \cite{deWit:1984nz,deWit:1986iy,Nicolai:2011cy,deWit:2013ija,Godazgar:2013nma,Godazgar:2013pfa} to four-dimensional solutions other than stationary points of the potential. It also shows both the power and usefulness of such formulae. 
It is also interesting to observe, once again, that while gauged $\Neql8$ supergravity is a very powerful tool in generating interesting new solutions, the solution we consider here appears  singular in four dimensions while its eleven-dimensional uplift is regular and far more physically interesting than the four-dimensional perspective would remotely suggest.  Once again, it illustrates the point noted long ago that to properly understand a holographic flow one must always look at it from the higher-dimensional perspective \cite{Warner:2000qh}.

There are many  questions that arise from this work and the first, and most obvious, is the extent it might be generalized.  There are some obvious first steps that are virtually guaranteed to work.   The $\rm SU(3)$ invariance of our solution required that the manifold $\cB_6$  be foliated by $\IC \IP_2$'s with the Fubini-Study metric.  Based on the experience coming from \cite{Corrado:2001nv}, we expect that $\IC \IP_2$ can easily be replaced by $S^2 \times S^2$ or any other K\"ahler-Einstein four-manifold.   In addition, the $\Neql2$ supergravity coupled to one vector multiplet that underlies our analysis can easily be generalized to $\Neql2$ supergravity coupled to three vector multiplets while still remaining within gauged $\Neql8$ supergravity.  The relevant truncation was considered in \cite{Freedman:2013oja} and the holomorphic superpotential replaces (\ref{holV1}) by
\begin{equation}
\mathcal{V} ~=~ \sqrt{2} (1+ z_1 z_2 z_3) \,,
\label{holV2}
\end{equation}
where the $z_i$ are the complex scalars of the three vector multiplets.  Our results here may be thought of as the special case with the three vector multiplets set equal and, in particular, $z_1=z_2=z_3=z$.  While the uplift formulae will be far more complicated, it is not difficult to guess that the infra-red limit will involve a more general K\"ahler manifold with a $\rm U(1)^3$ symmetry.  It may also have some non-trivial moduli in that the $\zeta = \pi/3$ condition may simply become a constraint on the overall phase of $z_1 z_2 z_3$.  These  moduli would probably be related to the three distinct sets of M5-brane fluxes on the K\"ahler manifold. We are investigating this further.
  
There is also the question of the infra-red limit as a separate compactification.   The metric and fluxes on  $\widehat {AdS_5} \times \cB_6$ are defined in a very canonical manner in terms of the K\"ahler structure on $\cB_6$.  Indeed our results are rather reminiscent of some of the conclusions of \cite{Gauntlett:2004zh} on the classification of  M-theory compactifications down to  $AdS_5 \times \cB_6$.  However, we have explicitly checked that the infra-red limit of our solutions {\it does not} by itself satisfy either the equations of motion or BPS conditions of supergravity:  It is simply not a separate and independent solution.  The infra-red limit of our flow is thus still inextricably linked to the flow that created it and some of the radially dependent terms contribute to the solving of the equations of motion in the infra-red limit.  In this respect the solution that we obtain here is similar to the experience in bubbling black holes:  while one starts with M2 brane charges, the infra-red is controlled by M5 brane fluxes but one simply cannot decouple the infra-red limit from the asymptotic solution in which it belongs.  

Having accepted that one must consider the whole flow and not the infra-red fixed point alone, it still raises the very interesting question as to whether one can make flows like this by replacing $S^7$ by any Sasaki-Einstein manifold, $\cM_7$.  While we suspect that such generalizations are indeed possible, they will have to be rather non-trivial.   One of the complications is that the base, $\cB_6$, does not have a K\"ahler form {\it along} our flow however $\cB_6$ {\it can} be thought of as a family of K\"ahler manifolds parametrized by the radial coordinate\footnote{The exterior derivative of the complex structure is always proportional to $dr$, where $r$ is the radial coordinate.}.  

On the other hand the observation of Section \ref{ss:CPW} came as rather a surprise: the K\"ahler metric on $\cB_6$ obtained here is {\it exactly} the same as the corresponding six-manifold that emerges at the conformal fixed point  of the flow in  \cite{Corrado:2001nv}.   This  suggests that the flows constructed here might be very special and the infra-red field theory on the intersecting M5 branes could be linked to the conformal fixed point in $2+1$ dimensions.  It is also possible that this ``K\"ahler  coincidence'' is   simply  an artefact of the high level of symmetry and the specialized truncation of $\Neql8$ supergravity.  It would be most interesting to seek generalizations of our new flow if only to shed light on the possible links between their infra-red limits and the conformal fixed point in $2+1$ dimensions.

Finally there are interesting questions relating to black-hole physics.  First, we have already noted that the dissolving of M2-brane charges into M5-brane fluxes is one of the core ingredients in the microstate geometry program for resolving black hole singularities and obviating the appearance of horizons \cite{Bena:2005va}.  Here we are seeing the same thing and indeed the observation just before (\ref{holV2}) shows that we really do have a ``three-charge'' family in which all three charges have been set equal, just as in the minimal ungauged supergravity relevant to black-hole physics.  Unlike the black-hole story, here the UV families of M2 branes involve only one species and yet there are potentially three sets of M5 branes in the infra-red.  In black hole physics, each pair of M5 branes intersect on a common $(3+1)$-dimensional space, but all three species only intersect in a  $(1+1)$-dimensional space.  In addition, the M5 branes emerging in the infra-red limit of the microstate geometries do not obviously produce a dielectric polarization of the supersymmetry but there are suggestions in which this may indeed be happening  on evanescent ergospheres \cite{Bena:2013ora}.  So we apparently have two rather different polarizations of M2 branes into M5 branes.  It would be extremely interesting to see if there is, in fact, a deeper relationship and use the results here to further inform black-hole physics, or vice versa. 
 
One might also examine a more direct relationship to black-hole physics by generalizing our solutions so as to include other charge sources.  One could start with some of the possible black-hole solutions in $AdS_4$  and turn on the scalar field considered here.  There would probably be a very interesting competition in the infra-red between our scalar and the formation of a black hole.  One might even find something akin to an $AdS_5$ black hole in the infra-red of what looks like an $AdS_4$ black hole in the UV.

It is evident that the new class of flows described here lead to a host of interesting and open problems.

\section*{Acknowledgments}

We would like to thank Ibrahima Bah, Iosif Bena and Nikolay Bobev for helpful discussions and comments on an early draft of the manuscript.
 This work was supported in part by  DOE~grant DE-SC0011687.





\begin{thebibliography}{99}


\bibitem{Myers:1999ps} 
  R.~C.~Myers,
``Dielectric branes,''
  JHEP {\bf 9912}, 022 (1999)
  [hep-th/9910053].
  
  
\bibitem{Polchinski:2000uf} 
  J.~Polchinski and M.~J.~Strassler,
  ``The String dual of a confining four-dimensional gauge theory,''
  hep-th/0003136.
  


\bibitem{Bena:2008dw} 
  I.~Bena, N.~Bobev, C.~Ruef and N.~P.~Warner,
``Supertubes in Bubbling Backgrounds: Born-Infeld Meets Supergravity,''
  JHEP {\bf 0907}, 106 (2009)
  [arXiv:0812.2942 [hep-th]].
  

\bibitem{Bena:2005va} 
  I.~Bena and N.~P.~Warner,
``Bubbling supertubes and foaming black holes,''
  Phys.\ Rev.\ D {\bf 74}, 066001 (2006)
  [hep-th/0505166].
 
\bibitem{Berglund:2005vb} 
  P.~Berglund, E.~G.~Gimon and T.~S.~Levi,
  ``Supergravity microstates for BPS black holes and black rings,''
  JHEP {\bf 0606}, 007 (2006)
  [hep-th/0505167].


\bibitem{Bena:2007kg} 
  I.~Bena and N.~P.~Warner,
  ``Black holes, black rings and their microstates,''
  Lect.\ Notes Phys.\  {\bf 755}, 1 (2008)
  [hep-th/0701216].
 
\bibitem{Ahn:2000mf} 
  C.~h.~Ahn and K.~Woo,
  ``Supersymmetric domain wall and RG flow from 4-dimensional gauged N=8 supergravity,''
  Nucl.\ Phys.\ B {\bf 599}, 83 (2001)
  [hep-th/0011121].
  
\bibitem{Ahn:2001by} 
  C.~h.~Ahn and K.~s.~Woo,
  ``Domain wall and membrane flow from other gauged d = 4, N=8 supergravity. Part 1,''
  Nucl.\ Phys.\ B {\bf 634}, 141 (2002)
  [hep-th/0109010].
  
\bibitem{Ahn:2001kw} 
  C.~h.~Ahn and T.~Itoh,
``An N = 1 supersymmetric G-2 invariant flow in M theory,''
  Nucl.\ Phys.\ B {\bf 627}, 45 (2002)
  [hep-th/0112010].
  
\bibitem{Ahn:2002eh} 
  C.~h.~Ahn and T.~Itoh,
  ``The Eleven-dimensional metric for AdS / CFT RG flows with common SU(3) invariance,''
  Nucl.\ Phys.\ B {\bf 646}, 257 (2002)
  [hep-th/0208137].
  
\bibitem{Ahn:2002qga} 
  C.~h.~Ahn and K.~s.~Woo,
  ``Domain wall from gauged d = 4, N=8 supergravity. Part 2,''
  JHEP {\bf 0311}, 014 (2003)
  [hep-th/0209128].
  
  
\bibitem{Bobev:2009ms} 
  N.~Bobev, N.~Halmagyi, K.~Pilch and N.~P.~Warner,
``Holographic, N=1 Supersymmetric RG Flows on M2 Branes,''
  JHEP {\bf 0909}, 043 (2009)
  [arXiv:0901.2736 [hep-th]].
  

\bibitem{Bobev:2010ib} 
  N.~Bobev, N.~Halmagyi, K.~Pilch and N.~P.~Warner,
``Supergravity Instabilities of Non-Supersymmetric Quantum Critical Points,''
  Class.\ Quant.\ Grav.\  {\bf 27}, 235013 (2010)
  [arXiv:1006.2546 [hep-th]].

  
  

\bibitem{Warner:1983du} 
  N.~P.~Warner,
 ``Some Properties of the Scalar Potential in Gauged Supergravity Theories,''
  Nucl.\ Phys.\ B {\bf 231}, 250 (1984).
  
\bibitem{Warner:1983vz} 
  N.~P.~Warner,
``Some New Extrema of the Scalar Potential of Gauged $N=8$ Supergravity,''
  Phys.\ Lett.\ B {\bf 128}, 169 (1983).
  
\bibitem{Corrado:2001nv} 
  R.~Corrado, K.~Pilch and N.~P.~Warner,
``An N=2 supersymmetric membrane flow,''
  Nucl.\ Phys.\ B {\bf 629}, 74 (2002)
  [hep-th/0107220].

\bibitem{Benna:2008zy} 
  M.~Benna, I.~Klebanov, T.~Klose and M.~Smedback,
``Superconformal Chern-Simons Theories and AdS(4)/CFT(3) Correspondence,''
  JHEP {\bf 0809}, 072 (2008)
  [arXiv:0806.1519 [hep-th]].
  
  

\bibitem{Cvetic:1999xx} 
  M.~Cvetic, S.~S.~Gubser, H.~Lu and C.~N.~Pope,
``Symmetric potentials of gauged supergravities in diverse dimensions and Coulomb branch of gauge theories,''
  Phys.\ Rev.\ D {\bf 62}, 086003 (2000)
  [hep-th/9909121].

\bibitem{Cvetic:2000zu} 
  M.~Cvetic, H.~Lu and C.~N.~Pope,
  ``Consistent sphere reductions and universality of the Coulomb branch in the domain wall / QFT correspondence,''
  Nucl.\ Phys.\ B {\bf 590}, 213 (2000)
  [hep-th/0004201].
  
  
\bibitem{Gowdigere:2002uk} 
  C.~N.~Gowdigere and N.~P.~Warner,
  ``Flowing with eight supersymmetries in M theory and F theory,''
  JHEP {\bf 0312}, 048 (2003)
  [hep-th/0212190].
  
  
\bibitem{Kraus:1998hv} 
  P.~Kraus, F.~Larsen and S.~P.~Trivedi,
  ``The Coulomb branch of gauge theory from rotating branes,''
  JHEP {\bf 9903}, 003 (1999)
  [hep-th/9811120].
  
\bibitem{Freedman:1999gk} 
  D.~Z.~Freedman, S.~S.~Gubser, K.~Pilch and N.~P.~Warner,
``Continuous distributions of D3-branes and gauged supergravity,''
  JHEP {\bf 0007}, 038 (2000)
  [hep-th/9906194].
  
\bibitem{Gowdigere:2005wq} 
  C.~N.~Gowdigere and N.~P.~Warner,
  ``Holographic Coulomb branch flows with N=1 supersymmetry,''
  JHEP {\bf 0603}, 049 (2006)
  [hep-th/0505019].
  
\bibitem{Pope:2003jp} 
  C.~N.~Pope and N.~P.~Warner,
``A Dielectric flow solution with maximal supersymmetry,''
  JHEP {\bf 0404}, 011 (2004)
  [hep-th/0304132].
  
\bibitem{Bena:2004jw} 
  I.~Bena and N.~P.~Warner,
  ``A Harmonic family of dielectric flow solutions with maximal supersymmetry,''
  JHEP {\bf 0412}, 021 (2004)
  [hep-th/0406145].


  
\bibitem{Lin:2004nb} 
  H.~Lin, O.~Lunin and J.~M.~Maldacena,
``Bubbling AdS space and 1/2 BPS geometries,''
  JHEP {\bf 0410}, 025 (2004)
  [hep-th/0409174].
  
  
\bibitem{Bobev:2013yra} 
  N.~Bobev, K.~Pilch and N.~P.~Warner,
``Supersymmetric Janus Solutions in Four Dimensions,''
  JHEP {\bf 1406}, 058 (2014)
  [arXiv:1311.4883 [hep-th]].

\bibitem{deWit:1984nz} 
  B.~de Wit, H.~Nicolai and N.~P.~Warner,
``The Embedding of Gauged $N=8$ Supergravity into $d=11$ Supergravity,''
  Nucl.\ Phys.\ B {\bf 255}, 29 (1985).
   
\bibitem{deWit:1986mz} 
  B.~de Wit and H.~Nicolai,
  ``$d=11$ Supergravity With Local SU(8) Invariance,''
  Nucl.\ Phys.\ B {\bf 274}, 363 (1986).

\bibitem{deWit:1986iy}
  B.~de Wit and H.~Nicolai,
  ``The Consistency of the S**7 Truncation in D=11 Supergravity,''
  Nucl.\ Phys.\ B {\bf 281} (1987) 211.
   
\bibitem{Nicolai:2011cy} 
  H.~Nicolai and K.~Pilch,
``Consistent Truncation of d = 11 Supergravity on AdS$_4 \times S^7$,''
  JHEP {\bf 1203}, 099 (2012)
  [arXiv:1112.6131 [hep-th]].
  
\bibitem{deWit:2013ija} 
  B.~de Wit and H.~Nicolai,
   ``Deformations of gauged SO(8) supergravity and supergravity in eleven dimensions,''
  JHEP {\bf 1305}, 077 (2013)
  [arXiv:1302.6219 [hep-th]].

\bibitem{Godazgar:2013nma} 
  H.~Godazgar, M.~Godazgar and H.~Nicolai,
   ``Testing the non-linear flux ansatz for maximal supergravity,''
  Phys.\ Rev.\ D {\bf 87}, 085038 (2013)
  [arXiv:1303.1013 [hep-th]].

\bibitem{Godazgar:2013pfa} 
  H.~Godazgar, M.~Godazgar and H.~Nicolai,
``Nonlinear Kaluza-Klein theory for dual fields,''
  Phys.\ Rev.\ D {\bf 88}, no. 12, 125002 (2013)
  [arXiv:1309.0266 [hep-th]].

 \bibitem{PTWnext} 
   K.~Pilch, A.~Tyukov and N.P.~Warner, to appear.

\bibitem{Maldacena:2000mw} 
  J.~M.~Maldacena and C.~Nunez,
  ``Supergravity description of field theories on curved manifolds and a no go theorem,''
  Int.\ J.\ Mod.\ Phys.\ A {\bf 16}, 822 (2001)
  [hep-th/0007018].
  
\bibitem{Gaiotto:2009gz} 
  D.~Gaiotto and J.~Maldacena,
``The Gravity duals of N=2 superconformal field theories,''
  JHEP {\bf 1210}, 189 (2012)
  [arXiv:0904.4466 [hep-th]].
  
\bibitem{Witten:1995ex} 
  E.~Witten,
   ``String theory dynamics in various dimensions,''
  Nucl.\ Phys.\ B {\bf 443}, 85 (1995)
  [hep-th/9503124].
  
  
  
\bibitem{Strominger:1995ac} 
  A.~Strominger,
  ``Open p-branes,''
  Phys.\ Lett.\ B {\bf 383}, 44 (1996)
  [hep-th/9512059].
  
\bibitem{Ganor:1996mu} 
  O.~J.~Ganor and A.~Hanany,
  ``Small E(8) instantons and tensionless noncritical strings,''
  Nucl.\ Phys.\ B {\bf 474}, 122 (1996)
  [hep-th/9602120].
  
\bibitem{Seiberg:1996vs} 
  N.~Seiberg and E.~Witten,
 ``Comments on string dynamics in six-dimensions,''
  Nucl.\ Phys.\ B {\bf 471}, 121 (1996)
  [hep-th/9603003].
  
\bibitem{Duff:1996cf} 
  M.~J.~Duff, H.~Lu and C.~N.~Pope,
  ``Heterotic phase transitions and singularities of the gauge dyonic string,''
  Phys.\ Lett.\ B {\bf 378}, 101 (1996)
  [hep-th/9603037].
  
\bibitem{Witten:1996qb} 
  E.~Witten,
  ``Phase transitions in M theory and F theory,''
  Nucl.\ Phys.\ B {\bf 471}, 195 (1996)
  [hep-th/9603150].
  
\bibitem{Klemm:1996bj}
  A.~Klemm, W.~Lerche, P.~Mayr, C.~Vafa and N.~P.~Warner,
``Selfdual strings and N=2 supersymmetric field theory,''
  Nucl.\ Phys.\ B {\bf 477} (1996) 746
  [hep-th/9604034].
  
\bibitem{Ganor:1996nf} 
  O.~J.~Ganor,
  ``Six-dimensional tensionless strings in the large N limit,''
  Nucl.\ Phys.\ B {\bf 489}, 95 (1997)
  [hep-th/9605201].

  
\bibitem{HoyosBadajoz:2010td} 
  C.~Hoyos,
  ``Higher dimensional conformal field theories in the Coulomb branch,''
  Phys.\ Lett.\ B {\bf 696}, 145 (2011)
  [arXiv:1010.4438 [hep-th]].
  
\bibitem{Young:2014jma} 
  D.~Young and K.~Zarembo,
  ``Holographic Dual of the Eguchi-Kawai Mechanism,''
  JHEP {\bf 1406}, 030 (2014)
  [arXiv:1404.0225 [hep-th]].
 



\bibitem{Aharony:2008ug} 
  O.~Aharony, O.~Bergman, D.~L.~Jafferis and J.~Maldacena,
 ``N=6 superconformal Chern-Simons-matter theories, M2-branes and their gravity duals,''
  JHEP {\bf 0810}, 091 (2008)
  [arXiv:0806.1218 [hep-th]].
  

\bibitem{Page:1985bq} 
  D.~N.~Page and C.~N.~Pope,
  ``Inhomogeneous Einstein Metrics On Complex Line Bundles,''
  Class.\ Quant.\ Grav.\  {\bf 4}, 213 (1987).
 
\bibitem{Gauntlett:2004hh} 
  J.~P.~Gauntlett, D.~Martelli, J.~F.~Sparks and D.~Waldram,
   ``A New infinite class of Sasaki-Einstein manifolds,''
  Adv.\ Theor.\ Math.\ Phys.\  {\bf 8}, 987 (2006)
  [hep-th/0403038].
 
\bibitem{Gowdigere:2003jf} 
  C.~N.~Gowdigere, D.~Nemeschansky and N.~P.~Warner,
 ``Supersymmetric solutions with fluxes from algebraic Killing spinors,''
  Adv.\ Theor.\ Math.\ Phys.\  {\bf 7}, 787 (2004)
  [hep-th/0306097].
  
\bibitem{Pilch:2004yg} 
  K.~Pilch and N.~P.~Warner,
``N = 1 supersymmetric solutions of IIB supergravity from Killing spinors,''
  hep-th/0403005.
  
\bibitem{Nemeschansky:2004yh} 
  D.~Nemeschansky and N.~P.~Warner,
``A Family of M theory flows with four supersymmetries,''
  hep-th/0403006.
  
 
  

 
  
\bibitem{Warner:2000qh} 
  N.~P.~Warner,
``Holographic renormalization group flows: the view from ten-dimensions,''
  Class.\ Quant.\ Grav.\  {\bf 18}, 3159 (2001)
  [hep-th/0011207].
  
\bibitem{Freedman:2013oja} 
  D.~Z.~Freedman and S.~S.~Pufu,
  ``The holography of $F$-maximization,''
  JHEP {\bf 1403}, 135 (2014)
  [arXiv:1302.7310 [hep-th]].

\bibitem{Gauntlett:2004zh} 
  J.~P.~Gauntlett, D.~Martelli, J.~Sparks and D.~Waldram,
``Supersymmetric AdS(5) solutions of M theory,''
  Class.\ Quant.\ Grav.\  {\bf 21}, 4335 (2004)
  [hep-th/0402153].


\bibitem{Bena:2013ora} 
  I.~Bena, S.~F.~Ross and N.~P.~Warner,
``On the Oscillation of Species,''
  JHEP {\bf 1409}, 113 (2014)
  [arXiv:1312.3635 [hep-th]].
  
\bibitem{Gubser:2000nd} 
  S.~S.~Gubser,
  ``Curvature singularities: The Good, the bad, and the naked,''
  Adv.\ Theor.\ Math.\ Phys.\  {\bf 4}, 679 (2000)
  [hep-th/0002160].
  

      
  
 \end{thebibliography}
\end{document}